\documentclass[aps,prb,twocolumn,superscriptaddress,showpacs,english]{revtex4-1}
\usepackage[T1]{fontenc}
\usepackage[latin9]{inputenc}
\usepackage{amsmath}
\usepackage{amssymb}
\usepackage{graphicx}
\usepackage{xcolor}
\usepackage{esint}
\definecolor{mydarkgreen}{rgb}{0.0,0.5,0.0}
\usepackage[linktocpage=true,
  colorlinks=true, 
  pdfborder={0 0 0},
  linkcolor=blue,
  citecolor=red,
  filecolor=yellow,
  urlcolor=mydarkgreen,
  bookmarks,
  pdftitle={paper},
  pdfauthor={},
]{hyperref}

\makeatletter

\usepackage{babel}
\usepackage{upgreek}

\makeatother

\begin{document}

\title{Hedin's Equations for Superconductors}

\author{A. Linscheid}
\affiliation{Max Planck Institute of Microstructure Physics, Weinberg 2, D-06120 Halle, Germany.}
\author{F. Essenberger}
\affiliation{Max Planck Institute of Microstructure Physics, Weinberg 2, D-06120 Halle, Germany.}

\begin{abstract}
We generalize Hedin's equations to a system of superconducting electrons
coupled with a system of phonons. The electrons are described by an
electronic Pauli Hamiltonian which includes the Coulomb interaction
among electrons and an external vector and scalar potential. We derive
the continuity equation in the presence of the superconducting condensate
and point out how to cast vertex corrections in the form of a non-local
effective interaction that can be used to describe both fluctuations
of spin and superconducting phase beyond the screened Coulomb self-energy
diagram.
\end{abstract}
\maketitle

\section{Introduction\label{sec:Introduction}}

Hedin's formally exact iterative procedure to generate the one particle
Green's function for electrons that interact via the Coulomb potential
\citep{HedinNewMethodForCalculatingTheOneParticleGF1965} has provided
a useful basis for further approximations. Especially the famous \textit{GW}
approximation, as already introduced in the original paper, is very
successful in the first principle analysis of physical system using
Many-Body-Perturbation theory \citep{OnidaElectronicExcitationsDFTVsMBGFApproaches}.
Nowadays, with increasing computational power, even a fully self-consistent
treatment of the screened Coulomb potential \textit{W} within Hedin's
cycle for both, solids and molecules is in reach \citep{SchoeneSelfConsistenGWinMetalsAndSemiconductors1998,CarusoSelfConsistenGWAllElectronMolecules2013}.

While we thus have a firm foundation of ab-initio electronic structure
perturbation theory, to calculate superconductivity (SC), the Green's
function approach boils down to the Eliashberg equations \citep{EliashbergInteractionBetweenElAndLatticeVibrInASC1960,ScalapinoStrongCouplSC1966,CarbottePropertiesOfBosonExchangeSC1990}.
There, the Coulomb potential is usually reduced into the $\mu^{\star}$
pseudo-potential \citep{CarbottePropertiesOfBosonExchangeSC1990,ScalapinoStrongCouplSC1966}
and the phonons are computed externally, i.e.~in the normal state.
This is reasonable since the vibrational structure and the renormalized
Coulomb potential appear to be largely independent on the SC condensation.
Thus, in the usual Eliashberg approach, SC is computed starting from
a converged normal state electronic band and phonon structure \citep{ScalapinoStrongCouplSC1966}.
On the other hand, spin-fluctuations are discussed as a pairing mechanism
in the unconventional SC such as the cuprates and iron pnictides \citep{manskeTheoryOfUnconventionalSC2004}.
With fluctuations we mean here self-energy corrections that describe
effects beyond the screened Coulomb diagram. The paramagnon peak as
the quasi-particle that corresponds to the spin-fluctuation interaction
becomes entirely different in the SC phase \citep{InosovNormalStateSpinDynamicsAndTempDependentSpinResonanceEnergyInOptimallyDopedBaFeCoAs2010}
and it therefore is essential to consider SC and the screening of
electronic interactions on the same footing. The paramagnetic spin-fluctuations
are included in ab-initio methods on a semi-ab-initio level
so far. Essenberger \textit{et al. \citep{EssenbergerSCPairingMediatedBySpinFluctuationsFromFirstPrinciples2015}}
have derived an effective interaction that is based on a self-energy
contribution due to the magnetic susceptibility and the exchange-correlation
kernel of time dependent density functional theory \citep{PetersilkaExcitationEnergiesFromTDDFT1996}.
We note, however, that the effective interaction derived in Ref.~\textit{\citep{EssenbergerSCPairingMediatedBySpinFluctuationsFromFirstPrinciples2015}}
is in the normal state and the self-energy that is used may suffer
from double counting. 
The diagrams considered for the construction of the
effective interaction are formally similar to approach of Berk and Schrieffer \citep{PhysRevLett.17.433}.
In the context of a Hubbard model, this approach has been extensivley used
to describe the SC phase of the iron based superconductors
(see Ref.~\onlinecite{HirschfeldReview} for a review).

In a similar manner fluctuations of the SC order parameter \citep{LarkinFluctuationPhenomenaInSuperconductors2008}
in constrained geometries are not well described in present ab-initio
methods. While the famous Mermin-Wagner theorem \citep{MerminAbsenceOfFerroOrAntiferroMagnetismIn1Or2D1966}
forbids the ordering in 2D due to the onset of long wavelength fluctuations
it is still unclear to what extend this result has implications in
experiment. For example, SC has been observed in a lead surface on
a silicone substrate down to the single atomic layer \citep{ZhangSuperconductivityInOneAtomicLayerMetalFilmsGrownOnSi111_2010}.
Similarly, taking the Coulomb interaction into account, the excitation
energies of the collective states of paired electrons starts at the
plasma frequency \citep{Anderson1958RPAInTheoryOfSuperconductivity,NambuQParticlesGaugeInSuperconductivity1960,Ambegaokar1961}.
Since plasmons with a small excitation energy may exist in low dimensional
systems, the collective excitations, i.e.~SC fluctuations, may have
to be taken into account self-consistently.

To present a fundamental theory to describe SC based on the basic
interactions among electrons is the content of this paper. In particular,
we derive Hedin's equations for SC electrons that interact via the
Coulomb potential and with a system of phonons in Sec.~\ref{sec:HedinEquations}.
Our approach treats the electronic screening, the coupling to the
phonons and SC on the same footing. In Sec.~\ref{sec:TheEffectiveInteraction},
we point how to interpret vertex corrections as an effective interaction,
without introducing double counting with respect to the screened Coulomb
interaction. Our coupled equations are capable to describe fluctuation
effects in principle exactly. This means in particular spin angular
momentum transfer processes in a paramagnetic system as well as, in
the language of Anderson \citep{Anderson1958RPAInTheoryOfSuperconductivity},
iso-spin momentum transfer with respect to the Nambu off diagonal
components, which corresponds to the Cooper-channel, of the Green's
function.

\section{The Hamiltonian}

In this Section we define the basic Hamiltonian governing our system
of electrons and phonons. Before we introduce the individual parts
of the Hamiltonian in Eq.~(\ref{eq:TotalHamiltonian}), we define
the notation that we use in this work in the following separate paragraph.

\paragraph{Notation}

To simplify the notation we introduce the usual thermal average $\langle...\rangle_{0}\equiv{\rm Tr}\bigl\{{\rm e}^{-\beta(\hat{H}-\mu\hat{N})}...\bigr\}/{\rm Tr}\bigl\{{\rm e}^{-\beta(\hat{H}-\mu\hat{N})}\bigr\}$
and $\langle\hat{A}\rangle_{{\rm {\scriptscriptstyle T}}}\equiv\langle{\rm T}\hat{U}(\beta,0)\hat{A}\rangle_{0}/\langle\hat{U}(\beta,0)\rangle_{0}$
with the time ordering symbol ${\rm T}$ and the time evolution operator
$\hat{U}(\tau,\tau^{\prime})$.

Following the notation of Nambu \citep{NambuQParticlesGaugeInSuperconductivity1960}
and Anderson \citep{Anderson1958RPAInTheoryOfSuperconductivity},
in order to describe it as a usual single particle Hamiltonian we
introduce the Nambu field via
\begin{eqnarray}
\mbox{\ensuremath{\hat{\varPsi}}}(\boldsymbol{{\it r}}) & = & \left(\begin{array}{cccc}
\hat{\psi}(\boldsymbol{{\it r}}\uparrow) & \hat{\psi}(\boldsymbol{{\it r}}\downarrow) & \hat{\psi}^{\dagger}(\boldsymbol{{\it r}}\uparrow) & \hat{\psi}^{\dagger}(\boldsymbol{{\it r}}\downarrow)\end{array}\right)^{{\rm T}_{{\scriptscriptstyle {\rm ns}}}}\,.
\end{eqnarray}
We use ${\rm T}_{{\scriptscriptstyle {\rm ns}}}$ to transpose in
Nambu (n) and spin space (s) and label Nambu components with $\alpha=1,-1$
and spin components with $\mu=\uparrow,\downarrow$. ${\rm Tr}_{{\rm {\scriptscriptstyle ns}}}$
means to take the trace in Nambu and spin space.

We further group variables in the notation
\begin{equation}
\begin{array}{cccccc}
{\it 1} & = & \boldsymbol{{\it r}}_{1}\tau_{1} & \quad\bar{{\it 1}} & = & \alpha_{1}\mu_{1}\\
\int\hspace{-0.1cm}{\rm d}{\it 1} & = & \int\hspace{-0.1cm}{\rm d}\tau_{1}\int\hspace{-0.1cm}{\rm d}\boldsymbol{{\it r}}_{1} & \quad\sum_{\bar{{\it 1}}} & = & \sum_{\mu_{1}\alpha_{1}}\\
\updelta_{1,2} & = & \updelta(\boldsymbol{{\it r}}_{1}-\boldsymbol{{\it r}}_{2})\times & \quad\updelta_{\bar{{\it 1}},\bar{{\it 2}}} & = & \updelta_{\alpha_{1}\alpha_{2}}\updelta_{\mu_{1}\mu_{2}}\\
 &  & \times\updelta(\tau_{1}-\tau_{2})
\end{array}
\end{equation}
The bar indicates \textit{Nambu-spin}. Thus, similarly, we indicate
matrices in Nambu and spin space with a bar on top. For example, in
this notation, the electronic single particle Green's function reads
\begin{eqnarray}
\bar{G}(\boldsymbol{{\it r}}\tau,\boldsymbol{{\it r}}^{\prime}\tau^{\prime}) & = & -\langle\hat{\varPsi}_{{\rm I}}(\boldsymbol{{\it r}}\tau)\otimes\hat{\varPsi}_{{\rm I}}^{\dagger}(\boldsymbol{{\it r}}^{\prime}\tau^{\prime})\rangle_{{\rm {\scriptscriptstyle T}}}\,.
\end{eqnarray}
The symbol $\otimes$ means to take the outer product in Nambu and
spin space and the subscript ${\rm I}$ refers to an operator in the
Heisenberg picture $\hat{O}_{{\rm I}}(\tau)=\hat{U}(0,\tau)\hat{O}\hat{U}(\tau,0)$.
We promote the Nambu and spin indices to the argument to refer to
the matrix elements, similar to the usual matrix notation. There,
for a matrix $A$, one refers to the matrix elements as $A_{ij}$.
$\bar{G}({\it 1}\bar{{\it 1}},{\it 2}\bar{{\it 2}})$ is thus the
$\alpha_{1}\sigma_{1},\alpha_{2}\sigma_{2}$ matrix element in Nambu
and spin space - a function of two space and time variables ${\it 1}$
and ${\it 2}$, respectively. We use this matrix element notation
also for higher order tensors, such as $I({\it 1}\bar{{\it 1}},{\it 2}\bar{{\it 2}},{\it 3}\bar{{\it 3}},{\it 4}\bar{{\it 4}})$
which is the $(\alpha_{1}\sigma_{1},\alpha_{2}\sigma_{2},\alpha_{3}\sigma_{3},\alpha_{4}\sigma_{4})$
component of the $4\times4\times4\times4$ object $I({\it 1},{\it 2},{\it 3},{\it 4})$.

If indexes appear on one side of an equation and not on the other
we define that these are summed or integrated.

We use the Pauli matrices both in spin as well as in Nambu space.
For the spin-Pauli matrices we use $\sigma_{0,x,y,z}$ while for the
latter we use $\tau_{0,x,y,z}$.

We introduce the Nambu time ordering symbol
\begin{eqnarray}
\bar{{\rm T}}\hat{\varPsi}_{{\rm I}}(\boldsymbol{{\it r}}\tau)\otimes\hat{\varPsi}_{{\rm I}}^{\dagger}(\boldsymbol{{\it r}}^{\prime}\tau^{\prime}) & = & \theta(\tau-\tau^{\prime})\hat{\varPsi}_{{\rm I}}(\boldsymbol{{\it r}}\tau)\otimes\hat{\varPsi}_{{\rm I}}^{\dagger}(\boldsymbol{{\it r}}^{\prime}\tau^{\prime})\nonumber \\
 & - & \theta(\tau^{\prime}-\tau)\bigl(\hat{\varPsi}_{{\rm I}}^{\dagger}(\boldsymbol{{\it r}}^{\prime}\tau^{\prime})\otimes\hat{\varPsi}_{{\rm I}}(\boldsymbol{{\it r}}\tau)\bigr)^{{\rm T}_{\text{sn}}}\nonumber \\
\label{eq:TimeOrderingSymbolNambu}
\end{eqnarray}
This form is equivalent to a time ordering in every component of the
Nambu and spin matrix. Note a peculiarity in the equal time limit:
We define that the $\alpha,\alpha^{\prime}=-1,-1$ component behaves
different to the $1,1$ component in the sense that here the second
argument is taken infinitesimally after the first. The reason is that
in the equal time limit it is necessary to recover the density operator
$\hat{n}(\boldsymbol{{\it r}})=\hat{\psi}^{\dagger}(\boldsymbol{{\it r}})\cdot\sigma_{0}\cdot\hat{\psi}(\boldsymbol{{\it r}})$,
also in the Nambu $-1,-1$ channel. This equal time limit appears
in the interactions in a diagrammatic expansion so the definition
must be chosen to recover this limit \citep{fetterWaleckaQuantumTheoryOfManyParticles1971}.
The missing reordering of the operators causes an additional minus
sign and thus $\hat{n}_{{\rm I}}(\boldsymbol{{\it r}}\tau)\equiv-(1/2){\rm Tr}_{{\rm {\scriptscriptstyle ns}}}\{\lim_{\tau^{\prime}\rightarrow\tau}\tau_{z}\cdot\bar{{\rm T}}\hat{\varPsi}_{{\rm I}}(\boldsymbol{{\it r}}\tau)\otimes\hat{\varPsi}_{{\rm I}}^{\dagger}(\boldsymbol{{\it r}}\tau^{\prime})\}$
with $\tau_{z}$ instead of $\tau_{0}$ without the reordering.

\paragraph{Contributions to the Hamiltonian}

Our starting point is a system of interacting electrons that is coupled
with a system of non-interacting phonons
\begin{eqnarray}
\hat{H} & = & \hat{H}_{{\scriptscriptstyle {\rm 0}}}+\hat{H}_{{\scriptscriptstyle {\rm p}}}+\hat{H}_{{\scriptscriptstyle {\rm e}-{\rm e}}}+\hat{H}_{{\scriptscriptstyle {\rm e}-{\rm p}}}+\hat{H}_{{\scriptscriptstyle {\rm aux}}}\,.\label{eq:TotalHamiltonian}
\end{eqnarray}
Here we distinguish the single particle part $\hat{H}_{{\scriptscriptstyle {\rm 0}}}$,
the non-interacting phonon part $\hat{H}_{{\scriptscriptstyle {\rm p}}}$,
the electron-electron interaction $\hat{H}_{{\scriptscriptstyle {\rm e}-{\rm e}}}$,
the electron-phonon interaction $\hat{H}_{{\scriptscriptstyle {\rm e}-{\rm p}}}$
and an auxiliary external potential that is set to zero after the
derivation. In the single particle part $\hat{H}_{{\scriptscriptstyle {\rm 0}}}$
we distinguish $\hat{H}_{{\scriptscriptstyle {\rm 0}}}=\hat{H}_{{\scriptscriptstyle {\rm n}}}+\hat{H}_{{\scriptscriptstyle {\rm s}}}$
with a normal state $\hat{H}_{{\scriptscriptstyle {\rm n}}}$ and
SC part $\hat{H}_{{\scriptscriptstyle {\rm s}}}$. The normal state
part is $\hat{H}_{{\scriptscriptstyle {\rm n}}}=\int{\rm d}\boldsymbol{{\it r}}\hat{\psi}^{\dagger}(\boldsymbol{{\it r}})\cdot\hat{H}_{{\scriptscriptstyle {\rm n}}}(\boldsymbol{{\it r}})\cdot\hat{\psi}(\boldsymbol{{\it r}})$
and $\hat{\psi}(\boldsymbol{{\it r}}\mu)$ is the electron spin $\mu$
field operator. Then, the normal state contribution of the single
particle part of the Hamiltonian reads 
\begin{eqnarray}
\hat{H}_{{\scriptscriptstyle {\rm n}}}(\boldsymbol{{\it r}}) & = & \sigma_{0}\Bigl(\frac{1}{2}\bigl(-{\rm i}\boldsymbol{\nabla}+\boldsymbol{A}_{{\scriptscriptstyle {\rm ext}}}(\boldsymbol{{\it r}})\bigr)^{2}+\phi_{{\scriptscriptstyle {\rm ext}}}(\boldsymbol{{\it r}})\Bigr)+\nonumber \\
 & + & \mathbf{S}\cdot\boldsymbol{B}_{{\scriptscriptstyle {\rm ext}}}(\boldsymbol{{\it r}})\,.\label{eq:NormalSingleParticleHamiltonian}
\end{eqnarray}
Here $\boldsymbol{B}_{{\scriptscriptstyle {\rm ext}}}(\boldsymbol{{\it r}})=\boldsymbol{\nabla}\times\boldsymbol{A}_{{\scriptscriptstyle {\rm ext}}}(\boldsymbol{{\it r}})$,
$\mathbf{S}=(\begin{array}{ccc}
\sigma_{x} & \sigma_{y} & \sigma_{z}\end{array})^{{\rm T}}$ and $\boldsymbol{A}_{{\scriptscriptstyle {\rm ext}}}(\boldsymbol{{\it r}})$
is an external vector potential. Further $\phi_{{\scriptscriptstyle {\rm ext}}}(\boldsymbol{{\it r}})$
is the scalar potential including the Coulomb attraction of the ions
in their equilibrium position. The existence of a pair condensate
with a macroscopic number of electron pairs is measured with the order
parameter of SC $\boldsymbol{\chi}(\boldsymbol{{\it r}},\boldsymbol{{\it r}}^{\prime})=\langle\hat{\psi}(\boldsymbol{{\it r}})\cdot\boldsymbol{\Phi}\cdot\hat{\psi}(\boldsymbol{{\it r}}^{\prime})\rangle$
where $\hat{\psi}(\boldsymbol{{\it r}})=\bigl(\begin{array}{cc}
\hat{\psi}(\boldsymbol{{\it r}}\uparrow) & \hat{\psi}(\boldsymbol{{\it r}}\downarrow)\end{array}\bigr)^{{\rm T}}$. The 4 component vector $\boldsymbol{\Phi}=(\begin{array}{cccc}
{\rm i}\sigma_{y} & -\sigma_{z} & \sigma_{0} & \sigma_{x}\end{array})^{{\rm T}}$ parametrizes the order parameter into 1 singlet $\Phi_{1}$ and 3
triplet parts $\Phi_{2},\Phi_{3},\Phi_{4}$. The singlet and triplet
SC part of the total Hamiltonian that couples to an external singlet
and triplet pairing field $\boldsymbol{\varDelta}_{{\scriptscriptstyle {\rm ext}}}^{\ast}(\boldsymbol{{\it r}},\boldsymbol{{\it r}}^{\prime})$
is
\begin{eqnarray}
\hat{H}_{{\scriptscriptstyle {\rm s}}} & = & -\frac{1}{2}\int\hspace{-0.2cm}{\rm d}\boldsymbol{{\it r}}\int\hspace{-0.2cm}{\rm d}\boldsymbol{{\it r}}^{\prime}\bigl(\hat{\psi}(\boldsymbol{{\it r}})\cdot\boldsymbol{\Phi}\cdot\hat{\psi}(\boldsymbol{{\it r}}^{\prime})\cdot\boldsymbol{\varDelta}_{{\scriptscriptstyle {\rm ext}}}^{\ast}(\boldsymbol{{\it r}},\boldsymbol{{\it r}}^{\prime})\nonumber \\
 & - & \hat{\psi}^{\dagger}(\boldsymbol{{\it r}})\cdot\boldsymbol{\Phi}\cdot\hat{\psi}^{\dagger}(\boldsymbol{{\it r}}^{\prime})\cdot\boldsymbol{\varDelta}_{{\scriptscriptstyle {\rm ext}}}(\boldsymbol{{\it r}},\boldsymbol{{\it r}}^{\prime})\bigr)\,.\label{eq:PairingHamiltonian}
\end{eqnarray}
Here, the combination $\boldsymbol{\varDelta}_{{\scriptscriptstyle {\rm ext}}}(\boldsymbol{{\it r}},\boldsymbol{{\it r}}^{\prime})\cdot\boldsymbol{\Phi}_{\mu\mu^{\prime}}=-\boldsymbol{\varDelta}_{{\scriptscriptstyle {\rm ext}}}(\boldsymbol{{\it r}}^{\prime},\boldsymbol{{\it r}})\cdot\boldsymbol{\Phi}_{\mu^{\prime}\mu}$
is totally antisymmetric. Thus, the pair Hamiltonian Eq.~\ref{eq:PairingHamiltonian}
is hermitian by construction. We use $q=\boldsymbol{{\it q}},\lambda$
with the Bloch vector $\boldsymbol{{\it q}}$ and the mode number
$\lambda$ to indicate the quantum number on the phonon operators
$\hat{b}_{q}$. Then, the phononic single particle Hamiltonian is
\begin{eqnarray}
\hat{H}_{{\scriptscriptstyle p}} & = & \sum_{q}\varOmega_{q}(\hat{b}_{q}^{\dagger}\hat{b}_{q}+\frac{1}{2})\,.
\end{eqnarray}
We assume the electrons to interact via a spontaneous Coulomb interaction
\begin{eqnarray}
\hat{H}_{{\scriptscriptstyle {\rm e}-{\rm e}}} & = & \frac{1}{2}\sum_{\mu\mu^{\prime}}\!\int\hspace{-0.2cm}{\rm d}\boldsymbol{{\it r}}\int\hspace{-0.2cm}{\rm d}\boldsymbol{{\it r}}^{\prime}w_{{\scriptscriptstyle {\rm 0}}}(\boldsymbol{{\it r}},\boldsymbol{{\it r}}^{\prime})\times\\
 & \times & \hat{\psi}^{\dagger}\!(\boldsymbol{{\it r}}\mu)\hat{\psi}^{\dagger}\!(\boldsymbol{{\it r}}^{\prime}\mu^{\prime})\hat{\psi}(\boldsymbol{{\it r}}^{\prime}\mu^{\prime})\hat{\psi}(\boldsymbol{{\it r}}\mu)\,,
\end{eqnarray}
with $w_{{\scriptscriptstyle {\rm 0}}}(\boldsymbol{{\it r}},\boldsymbol{{\it r}}^{\prime})=1/\vert\boldsymbol{{\it r}}-\boldsymbol{{\it r}}^{\prime}\vert$.
$w_{{\scriptscriptstyle {\rm 0}}}(\boldsymbol{{\it r}},\boldsymbol{{\it r}}^{\prime})$
is independent of spin and also the screened potential will share
this property. Following Ref.~\citep{AryasetiawanGeneralizedHedinEquationsSpinDependentInteractions2008}
one could start with a spin dependent interaction. In this work we
take a different route. After the derivation of Hedin's equations
for a SC, we discuss the important self-energy diagrams that form
an effective spin dependent interaction.and the electron phonon interaction
is (we use the notation $-q\equiv-\boldsymbol{{\it q}},\lambda$ and
$i=0,x,y,z$)
\begin{eqnarray}
\hat{H}_{{\scriptscriptstyle {\rm e}-{\rm p}}} & = & \sum_{iq}\int\hspace{-0.2cm}{\rm d}\boldsymbol{{\it r}}\hat{\psi}^{\dagger}(\boldsymbol{{\it r}})\cdot\sigma_{i}\cdot\hat{\psi}(\boldsymbol{{\it r}})g_{i}(\boldsymbol{{\it r}}q)(\hat{b}_{q}+\hat{b}_{-q}^{\dagger})\,.
\end{eqnarray}
Because $\hat{H}_{{\scriptscriptstyle {\rm e}-{\rm p}}}$ is hermitian,
$g_{i}(\boldsymbol{{\it r}}q)$ satisfies $g_{i}(\boldsymbol{{\it r}}q)=g_{i}^{\ast}(\boldsymbol{{\it r}},-q)$.
The operator $\hat{H}_{{\scriptscriptstyle {\rm s}}}$ is not diagonal
in the electronic field operator $\hat{\psi}(\boldsymbol{{\it r}})$.
To derive the Hedin equations we use the auxiliary external fields
$\varphi(\boldsymbol{r}\tau)$ and $J_{q}(\tau)$ that are set to
zero after the derivation. We define
\begin{eqnarray}
\hat{H}_{{\scriptscriptstyle {\rm aux}}} & = & \hat{H}_{{\scriptscriptstyle {\rm aux}}}^{0}+\hat{H}_{{\scriptscriptstyle {\rm aux}}}^{p}\,,
\end{eqnarray}
and choose $\hat{H}_{{\scriptscriptstyle {\rm aux}}}^{0}(\tau_{1})=\int{\rm d}\boldsymbol{r}_{1}\varphi({\it 1})\hat{\psi}^{\dagger}(\boldsymbol{{\it r}}_{1})\cdot\sigma_{0}\cdot\hat{\psi}(\boldsymbol{{\it r}}_{1})$
and $\hat{H}_{{\scriptscriptstyle {\rm aux}}}^{p}=J_{q}(\hat{b}_{q}+\hat{b}_{-q}^{\dagger})$
where $J_{-q}=J_{q}^{\ast}$ to make the operator hermitian.

\section{Hedin's Equations for Superconductors\label{sec:HedinEquations}}

In this Section, we derive the Hedin equations for a superconductor.
We point out that we can work in the ground state, the equilibrium
finite temperature \citep{fetterWaleckaQuantumTheoryOfManyParticles1971}
or the more general Keldysh formalism\textcolor{red}{{} \citep{StefanucciNonequilibriumMBTOfQuantumSystemsAModernIntroduction2013}}.
The Hedin equation formally have the same shape. Here we work in the
equilibrium finite temperature formalism for definiteness. Our starting
point is the equation of motion of the single particle Green's function.
We express the two particle part as a functional derivative with respect
to the auxiliary field. Using a similar strategy for the appearing
self-energy and vertex terms leads to a set of closed equations, independent
on the auxiliary potential - the Hedin equations.

\subsection{Green's Function Dyson Equation}

In the following we derive the Dyson equation for the electronic Green's
function from the equation of motion
\begin{eqnarray}
-\partial_{\tau_{1}}\bar{G}({\it 1}\bar{{\it 1}},{\it 2}\bar{{\it 2}}) & = & \updelta_{{\it 1},{\it 2}}\updelta_{\bar{{\it 1}},\bar{{\it 2}}}+\nonumber \\
 & + & \langle[\hat{H},\hat{\varPsi}(\boldsymbol{{\it r}}_{1}\bar{{\it 1}})]_{{\rm I}}(\tau_{1})\hat{\varPsi}_{{\rm I}}^{\dagger}({\it 2}\bar{{\it 2}})\rangle_{{\rm {\scriptscriptstyle T}}}\,.\label{eq:GFEquationOfMotion}
\end{eqnarray}
Similarly to the original derivation, we are describing the two particle
part of the equation of motion as the functional derivative with respect
to the auxiliary field $\varphi({\it 1})$. Note here that although
$\hat{H}$ and $\updelta\hat{H}/\updelta\varphi(1)$ do not commute
in the case of a SC, the time ordering still allows us to use $\frac{\updelta}{\updelta\varphi(1)}\bar{{\rm T}}\hat{U}(\beta,0)=\frac{1}{2}\bar{{\rm T}}\hat{U}(\beta,0)\hat{\varPsi}_{{\rm I}}^{\dagger}({\it 1})\cdot\sigma_{0}\tau_{z}\cdot\hat{\varPsi}_{{\rm I}}({\it 1})$.
We complete the derivation of the Dyson equation in the Appendix \ref{sec:The-Electronic-Dyson}
with the result
\begin{eqnarray}
\bar{G}({\it 1},{\it 2}) & = & \bar{G}_{{\rm {\scriptscriptstyle H}}}({\it 1},{\it 2})+\bar{G}_{{\rm {\scriptscriptstyle H}}}({\it 1},{\it 3})\cdot\bar{\varSigma}({\it 3},{\it 4})\cdot\bar{G}({\it 4},{\it 2})\,.\nonumber \\
\label{eq:GFDysonEquation}
\end{eqnarray}
Also the self-energy $\bar{\varSigma}({\it 3},{\it 4})$ is defined
in the Appendix \ref{sec:The-Electronic-Dyson}. Our Dyson equation
starts from the Hartree Green's function $\bar{G}_{{\rm {\scriptscriptstyle H}}}({\it {\it 3}},{\it 2})$
that satisfies
\begin{eqnarray}
\bigl(-\tau_{0}\sigma_{0}\partial_{\tau_{1}}\delta_{{\it 1},{\it 3}}-\hat{\bar{H}}_{{\rm {\scriptscriptstyle H}}}({\it 1},{\it 3})\bigr)\cdot\bar{G}_{{\rm {\scriptscriptstyle H}}}({\it {\it 3}},{\it 2}) & = & \updelta_{{\it 1},{\it 2}}\sigma_{0}\tau_{0}\nonumber \\
\label{eq:HartreeGreensfunction}
\end{eqnarray}
where the Nambu and spin matrix operator $\hat{\bar{H}}_{{\rm {\scriptscriptstyle H}}}({\it 1},{\it 3})$
in first quantization is given by $\hat{\bar{H}}_{{\rm {\scriptscriptstyle H}}}({\it 1},{\it 3})=\hat{\bar{H}}_{{\scriptscriptstyle 0}}({\it 1},{\it 3})+\varPhi({\it 1})\sigma_{0}\tau_{z}\delta_{{\it 1},{\it 3}}$
with the Hartree field $\varPhi({\it 1})=\int n(\boldsymbol{{\it r}})w_{{\scriptscriptstyle {\rm 0}}}(\boldsymbol{{\it r}}_{1},\boldsymbol{{\it r}}){\rm d}\boldsymbol{{\it r}}+\varphi({\it 1})$
where $n(\boldsymbol{r})$ is the electron density. The bare Hamiltonian
is 
\begin{eqnarray}
\hat{\bar{H}}_{{\scriptscriptstyle {\rm 0}}}({\it 1},{\it 3}) & = & \left(\begin{array}{cc}
\hat{H}_{{\scriptscriptstyle {\rm n}}}({\it 1})\updelta_{{\it 1,3}} & \boldsymbol{\varDelta}_{{\scriptscriptstyle {\rm ext}}}({\it 1},{\it 3})\cdot\boldsymbol{\Phi}\\
-\boldsymbol{\varDelta}_{{\scriptscriptstyle {\rm ext}}}^{\ast}({\it 1},{\it 3})\cdot\boldsymbol{\Phi} & -\bigl(\hat{H}_{{\scriptscriptstyle {\rm n}}}({\it 1})\bigr)^{{\rm T}_{{\scriptscriptstyle {\rm s}}}}\updelta_{{\it 1,3}}
\end{array}\right)\,,\nonumber \\
\end{eqnarray}
with the normal state part $\hat{H}_{{\scriptscriptstyle {\rm n}}}$
of the total Hamiltonian given in Eq.~(\ref{eq:NormalSingleParticleHamiltonian}).
Note that because of the total anti-symmetry of $\boldsymbol{\varDelta}_{{\scriptscriptstyle {\rm ext}}}({\it 1},{\it 3})\cdot\boldsymbol{\Phi}$
and $\mathbf{S}^{{\rm T}_{{\scriptscriptstyle {\rm s}}}}=\mathbf{S}^{\ast}$
the bare Hamiltonian $\hat{\bar{H}}_{{\scriptscriptstyle {\rm 0}}}$
is hermitian. The self-energy $\bar{\varSigma}({\it 3},{\it 4})$
so far remains in the form given in the Appendix \ref{sec:The-Electronic-Dyson},
Eq.~(\ref{eq:AppendixSelfEnergy}) which depends on the auxiliary
potentials. In the next two Subsections we give self consistent equations
for the phononic and electronic part in terms of the Green's function
and the vertex, thereby removing this dependence on the auxiliary
potentials.

\subsection{Phonon Propagator, Vertex and Self-Energy}

In this Subsection cast the functional derivative with respect to
the auxiliary potentials into a dependence on the Green's function
only. As a preparatory, we introduce the bare phonon vertex 
\begin{eqnarray}
\bar{\varGamma}_{{\scriptscriptstyle {\rm ph}}}^{{\scriptscriptstyle {\rm 0}}}({\it 1}\bar{{\it 1}},{\it 2}\bar{{\it 2}};q\tau_{3}) & = & (\bar{v}_{i})_{\bar{{\it 1}}\bar{{\it 2}}}g_{i}(\boldsymbol{{\it r}}_{1}q)\updelta_{{\it 1}{\it 2}}\updelta_{\tau_{3}\tau_{1}}\,,\label{eq:BarePhononVertex}
\end{eqnarray}
where $\bar{\boldsymbol{v}}=(\begin{array}{cccc}
\tau_{z}\sigma_{0} & \tau_{z}\sigma_{x} & \tau_{0}\sigma_{y} & \tau_{z}\sigma_{z}\end{array})$ and we use the abbreviation $\hat{a}_{q}=\hat{b}_{q}+\hat{b}_{-q}^{\dagger}$.
The self-energy $\bar{\varSigma}$ of Eq.~(\ref{eq:AppendixSelfEnergy})
in the Appendix \ref{sec:The-Electronic-Dyson} consists of the parts
$\bar{\varSigma}({\it 1},{\it 2})=\bar{\varSigma}_{{\scriptscriptstyle {\rm xc}}}({\it 1},{\it 2})+\bar{\varSigma}_{{\scriptscriptstyle {\rm ph}}}^{{\scriptscriptstyle {\rm H}}}({\it 1},{\it 2})$.
Here, $\bar{\varSigma}_{{\scriptscriptstyle {\rm xc}}}({\it 1},{\it 2})=\bar{\varSigma}_{{\scriptscriptstyle {\rm C}}}({\it 1},{\it 2})+\bar{\varSigma}_{{\scriptscriptstyle {\rm ph}}}({\it 1},{\it 2})$
which separates an electronic and a phononic part. Thus, phonon self-energy
contributions to $\bar{\varSigma}({\it 1},{\it 2})$ are given by
\begin{eqnarray}
\bar{\varSigma}_{{\scriptscriptstyle {\rm ph}}}^{{\scriptscriptstyle {\rm H}}}({\it 1},{\it 2}) & = & \bar{\varGamma}_{{\scriptscriptstyle {\rm ph}}}^{{\scriptscriptstyle {\rm 0}}}({\it 1},{\it 2};q\tau_{3})\langle\hat{a}_{q}(\tau_{3})\rangle_{{\rm {\scriptscriptstyle T}}}\label{eq:PhononHartreeSE}\\
\bar{\varSigma}_{{\scriptscriptstyle {\rm ph}}}({\it 1},{\it 2}) & = & \bar{\varGamma}_{{\scriptscriptstyle {\rm ph}}}^{{\scriptscriptstyle {\rm 0}}}({\it 1},{\it 3};q\tau_{4})\cdot\bar{G}({\it 3},{\it 4})\cdot\nonumber \\
 &  & \cdot\bar{\varGamma}_{{\scriptscriptstyle {\rm ph}}}({\it 4},{\it 2};q^{\prime}\tau^{\prime})D_{{\scriptscriptstyle {\rm ph}}}(q\tau_{4},-q^{\prime}\tau_{5})\,.\label{eq:PhononSE}
\end{eqnarray}
The Hartree part $\bar{\varSigma}_{{\scriptscriptstyle {\rm ph}}}^{{\scriptscriptstyle {\rm H}}}({\it 1},{\it 2})$
is separated in order that we obtain a phononic vertex equation with
a bare term Eq.~(\ref{eq:BarePhononVertex}), later. We have introduced
the phonon propagator $D_{{\scriptscriptstyle {\rm ph}}}(q\tau,q^{\prime}\tau^{\prime})=\updelta\langle\hat{a}_{-q^{\prime}}(\tau^{\prime})\rangle_{{\rm {\scriptscriptstyle T}}}/\updelta J_{q}(\tau)$
and the phonon vertex $\bar{\varGamma}_{{\scriptscriptstyle {\rm ph}}}({\it 1},{\it 2};q\tau)=\updelta\bar{G}^{-1}({\it 1},{\it 2})/\updelta\langle\hat{a}_{q}(\tau)\rangle_{{\rm {\scriptscriptstyle T}}}$.
Evaluating commutators with the Hamiltonian Eq.~(\ref{eq:TotalHamiltonian})
and using $\varOmega_{q}=\varOmega_{-q}$ we obtain the equation of
motion
\begin{eqnarray}
T\partial_{\tau}^{2}\hat{a}_{q}(\tau) & = & 2\varOmega_{q}^{2}\hat{a}_{q}(\tau)\nonumber \\
 & - & 2\varOmega_{q}\bar{\varGamma}_{{\scriptscriptstyle {\rm ph}}}^{{\scriptscriptstyle {\rm 0}}}({\it 1}\bar{{\it 1}},{\it 2}\bar{{\it 2}};-q\tau_{3})T\hat{\varPsi}({\it 1}\bar{{\it 1}})\hat{\varPsi}^{\dagger}({\it 2}\bar{{\it 2}})\,.\nonumber \\
\end{eqnarray}
Further using Eq.~(\ref{eq:AppendixDerivative3}) it is straight
forward to derive a Dyson equation for $D_{{\scriptscriptstyle {\rm ph}}}(q\tau,q^{\prime}\tau^{\prime})$
\begin{eqnarray}
D_{{\scriptscriptstyle {\rm ph}}}(q\tau,q^{\prime}\tau^{\prime}) & = & D_{{\scriptscriptstyle {\rm ph}}}^{{\rm {\scriptscriptstyle 0}}}(q\tau,q^{\prime}\tau^{\prime})+D_{{\scriptscriptstyle {\rm ph}}}^{{\rm {\scriptscriptstyle 0}}}(q\tau,k\tau_{k})\times\nonumber \\
 & \times & \bar{\varGamma}_{{\scriptscriptstyle {\rm ph}}}^{{\scriptscriptstyle {\rm 0}}}({\it 1}\bar{{\it 1}},{\it 2}\bar{{\it 2}};-q\tau_{3})\bar{G}({\it 3}\bar{{\it 3}},{\it 2}\bar{{\it 2}})\times\nonumber \\
 & \times & \bar{G}({\it 1}\bar{{\it 1}},{\it 4}\bar{{\it 4}})\bar{\varGamma}_{{\scriptscriptstyle {\rm ph}}}({\it 4}\bar{{\it 4}},{\it 3}\bar{{\it 3}};k^{\prime}\tau_{k}^{\prime})\times\nonumber \\
 & \times & D_{{\scriptscriptstyle {\rm ph}}}(k^{\prime}\tau_{k}^{\prime},q^{\prime}\tau^{\prime})\,,\label{eq:PhononDysonEquation}
\end{eqnarray}
with the bare phonon propagator
\begin{equation}
\bigl(\partial_{\tau}^{2}-\varOmega_{q}^{2}\bigr)D_{{\scriptscriptstyle {\rm ph}}}^{{\rm {\scriptscriptstyle 0}}}(q\tau,q^{\prime}\tau^{\prime})=2\varOmega_{q}\updelta_{q,-q^{\prime}}\updelta(\tau-\tau^{\prime})\,.
\end{equation}
The phonon vertex in turn satisfies
\begin{eqnarray}
\bar{\varGamma}_{{\scriptscriptstyle {\rm ph}}}({\it 1},{\it 2};q^{\prime}\tau^{\prime}) & = & \bar{\varGamma}_{{\scriptscriptstyle {\rm ph}}}^{{\scriptscriptstyle {\rm 0}}}({\it 1},{\it 2};q\tau_{3})+\frac{\updelta\bar{\varSigma}_{{\scriptscriptstyle {\rm xc}}}({\it 1}\bar{{\it 1}},{\it 2}\bar{{\it 2}})}{\updelta\bar{G}({\it 4}\bar{{\it 4}},{\it 5}\bar{{\it 5}})}\times\nonumber \\
 & \times & \bar{G}({\it 4}\bar{{\it 4}},{\it 6}\bar{{\it 6}})\bar{\varGamma}_{{\scriptscriptstyle {\rm ph}}}({\it 6}\bar{{\it 6}},{\it 7}\bar{{\it 7}};q\tau_{3})\bar{G}({\it 7}\bar{{\it 7}},{\it 5}{\it \bar{5}})\,.\nonumber \\
\label{eq:PhononVertex}
\end{eqnarray}
Thus all terms in the Eqs.~(\ref{eq:PhononHartreeSE}) and (\ref{eq:PhononSE})
do not dependent explicitly on the auxiliary potentials.

\subsection{Coulomb Vertex and Self-Energy}

\begin{figure}
\includegraphics[width=1\columnwidth]{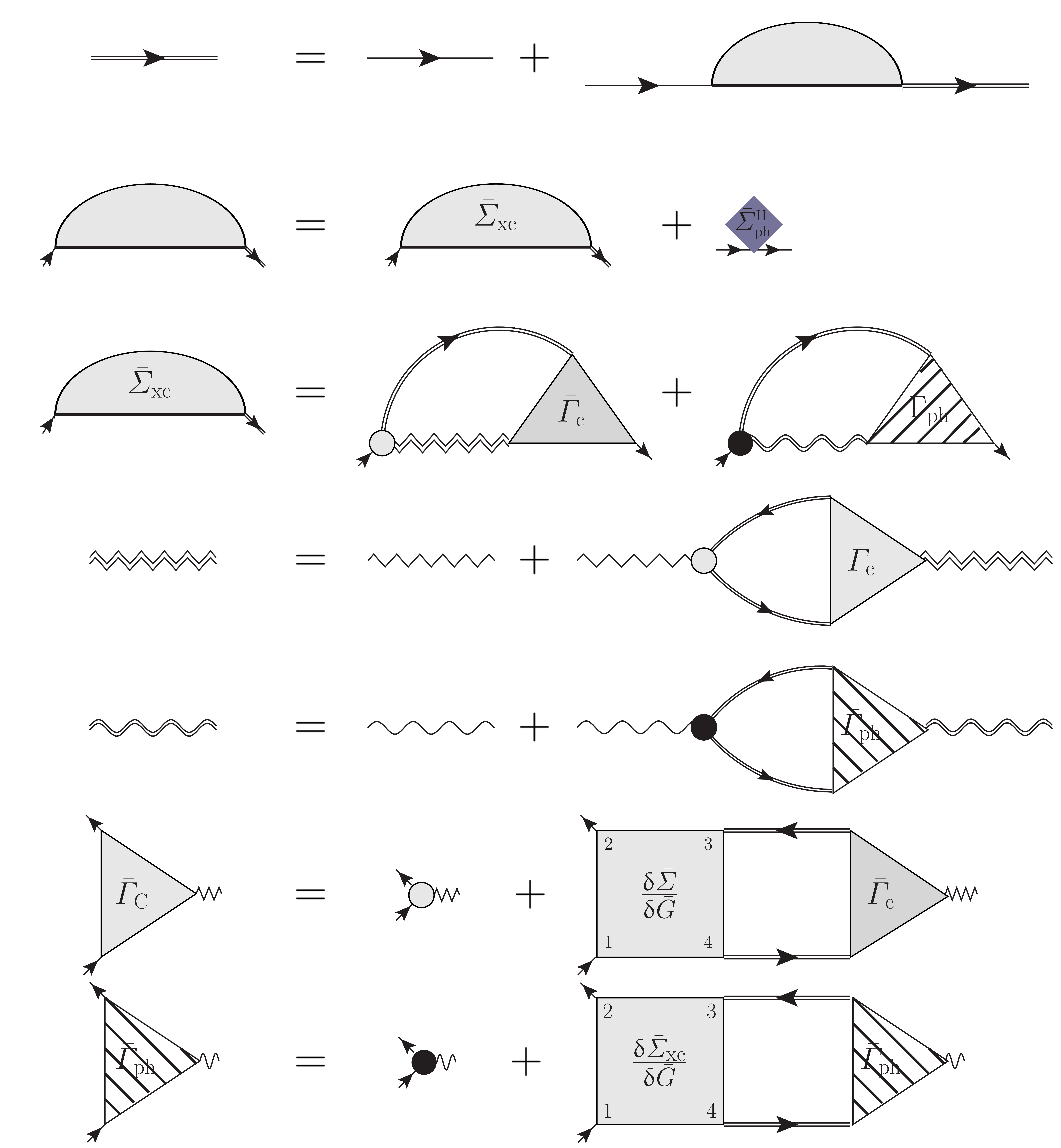}

\caption{Diagrammatic representation of the Hedin equations for a superconductors.\label{fig:DiagramaticHedin}}
\end{figure}

Repeating the above procedure for the electron-electron interaction
leads to an expression for the electronic self-energy in terms of
the electron vertex and the screened interaction. Using the chain
rule, the electronic self-energy in Eq.~(\ref{eq:AppendixSelfEnergy})
can be cast into
\begin{eqnarray}
\bar{\varSigma}_{{\scriptscriptstyle {\rm C}}}({\it 1},{\it 2}) & = & w({\it 1},{\it 4})\tau_{z}\sigma_{0}\cdot\bar{G}({\it 1},{\it 3})\cdot\bar{\varGamma}_{{\scriptscriptstyle {\rm C}}}({\it 3},{\it 2};{\it 4})\,.\label{eq:ElectronicSelfEnergy}
\end{eqnarray}
We introduce $\bar{\varGamma}_{{\scriptscriptstyle {\rm C}}}({\it 1},{\it 2};{\it 3})=-\updelta\bar{G}^{-1}({\it 1},{\it 2})/\updelta\varPhi({\it 3})$
as the electronic vertex and derive its Dyson equation with the notation
$\bar{\varGamma}_{{\scriptscriptstyle {\rm C}}}^{{\scriptscriptstyle {\rm 0}}}({\it 1},{\it 2};{\it 3})=\sigma_{0}\tau_{z}\delta_{{\it 1},{\it 2}}\delta_{{\it 1},{\it 3}}$
\begin{eqnarray}
\bar{\varGamma}_{{\scriptscriptstyle {\rm C}}}({\it 1},{\it 2};{\it 3}) & = & \bar{\varGamma}_{{\scriptscriptstyle {\rm C}}}^{{\scriptscriptstyle {\rm 0}}}({\it 1},{\it 2};{\it 3})+\frac{\updelta\bar{\varSigma}({\it 1}\bar{{\it 1}},{\it 2}\bar{{\it 2}})}{\updelta\bar{G}({\it 4}\bar{{\it 4}},{\it 5}\bar{{\it 5}})}\times\nonumber \\
 & \times & \bar{G}({\it 4}\bar{{\it 4}},{\it 6}\bar{{\it 6}})\bar{\varGamma}_{{\scriptscriptstyle {\rm C}}}({\it 6}\bar{{\it 6}},{\it 7}\bar{{\it 7}};{\it 3})\bar{G}({\it 7}\bar{{\it 7}},{\it 5}\bar{{\it 5}})\,.\nonumber \\
\label{eq:CoulombVertex}
\end{eqnarray}
At this point we have a Dyson equation for the electronic single particle
Green function (Eq.~(\ref{eq:GFDysonEquation})), the phonon propagator
(Eq.~(\ref{eq:PhononDysonEquation})) and the vertices (Eqs.~(\ref{eq:PhononVertex})
and (\ref{eq:CoulombVertex})) to construct the self-energy self-consistently.
The self-energy $\bar{\varSigma}_{{\scriptscriptstyle {\rm C}}}$
contains, in addition, the screened interaction $w({\it 1},{\it 2})=w_{{\scriptscriptstyle {\rm 0}}}({\it 1},{\it 3})\updelta\varPhi({\it 3})/\updelta\varphi({\it 2})$.
Inserting the definition of the Hartree field and using the chain
rule we conclude that $w({\it 1},{\it 2})$ satisfies 
\begin{eqnarray}
w({\it 1},{\it 2}) & = & w_{{\scriptscriptstyle {\rm 0}}}({\it 1},{\it 2})+w_{{\scriptscriptstyle {\rm 0}}}({\it 1},{\it 3})P({\it 3},{\it 4})w({\it 2},{\it 4})\label{eq:screendDynamicCoulomb}
\end{eqnarray}
where the polarization $P({\it 1},{\it 2})=\updelta n({\it 1})/\updelta\varPhi({\it 2})$
can be cast into
\begin{eqnarray}
P({\it 1},{\it 2}) & = & \frac{1}{2}{\rm Tr}_{{\rm {\scriptscriptstyle ns}}}\bigl\{\sigma_{0}\tau_{z}\cdot\bar{G}({\it 1},{\it 3})\cdot\bar{\varGamma}_{{\scriptscriptstyle {\rm C}}}({\it 3},{\it 4};{\it 2})\cdot\bar{G}({\it 4},{\it 1})\bigr\}\,,\nonumber \\
\label{eq:PolarizationEquation}
\end{eqnarray}
where we have used $n({\it 1})={\rm Tr}_{{\rm {\scriptscriptstyle ns}}}\bigl\{\sigma_{0}\tau_{z}\cdot\bar{G}({\it 1},{\it 1})\bigr\}/2$
which concludes our self-consistent Hedin cycle. This collection of
self-consistent equations is represented diagrammatically in Fig.~\ref{fig:DiagramaticHedin}
. In $P({\it 1},{\it 2})$, there appear contributions from the Cooper
channel that add to the screening of the bare Coulomb interaction
according to Eq.~\ref{eq:screendDynamicCoulomb}. For example, inserting
the bare Coulomb vertex $\bar{\varGamma}_{{\scriptscriptstyle {\rm C}}}^{{\scriptscriptstyle {\rm 0}}}$
for $\bar{\varGamma}_{{\scriptscriptstyle {\rm C}}}$ in Eq.~\ref{eq:PolarizationEquation}
the term $-\frac{1}{2}\int{\rm d}{\it 3}{\rm Tr}_{{\rm {\scriptscriptstyle s}}}\bigl\{\bar{G}({\it 1},-1,{\it 3},1)\bar{G}({\it 3},1,{\it 1},-1)\bigr\}$
with the SC Nambu off diagonal parts of the Green's function contributes
to the polarization and thus directly to the screening.

In practice, often $g_{i}(\boldsymbol{{\it r}}q)$ is taken as $\sum_{\mu\mu^{\prime}}\sigma_{i}^{\mu\mu^{\prime}}\updelta v_{{\scriptscriptstyle {\rm scf}}}(\boldsymbol{{\it r}},\mu\mu^{\prime})/\updelta u_{q}$
of a phonon calculation in density functional perturbation theory
(DFPT) and similarly the phonon mode frequencies are calculated from
the density response of a Kohn-Sham system in density functional theory
\citep{BaroniPhononrev2001}. The $\varOmega_{q}$ obtained in this
way include electronic screening already and often compare well with
experiment. In this case, care has to be taken not to double count
electronic screening, when Eq.~(\ref{eq:PhononDysonEquation}) is
applied \citep{vanLeeuwenFirstPrinciplesApproachToTheElPhInteraction2004}.

The question may be reformulated to what phonon system with frequencies
$\varOmega_{q}$ as ``bare'' we should start from in order that
a coupling to the electrons via Eq.~(\ref{eq:PhononDysonEquation})
leads to the best dressed phonon frequencies. While for calculating
SC it is often enough to neglect the influence of the electronic system
on the phonon system beyond DFPT, it would be clearly interesting
to investigate the influence of the SC phase on the phonon frequencies.
This issue must be regarded as an open problem as of now and a further
discussion is beyond the scope of this paper. In the following we
shall rely on the usual approach to consider the phonons as bare $D_{{\scriptscriptstyle {\rm ph}}}\rightarrow D_{{\scriptscriptstyle {\rm ph}}}^{{\rm {\scriptscriptstyle 0}}}$
and furthermore we follow an argument of Migdal \citep{MigdalInteractionBetweenElAndLatticeVibrInANormalMetal1958}
and Eliashberg \citep{EliashbergInteractionBetweenElAndLatticeVibrInASC1960}
and do not consider vertex corrections, i.e.~$\bar{\varGamma}_{{\scriptscriptstyle {\rm ph}}}\rightarrow\bar{\varGamma}_{{\scriptscriptstyle {\rm ph}}}^{{\scriptscriptstyle {\rm 0}}}$.

\section{The continuity equation\label{sec:The-continuity-equation}}

The external pairing field $\boldsymbol{\varDelta}_{{\scriptscriptstyle {\rm ext}}}$
and $\boldsymbol{\varDelta}_{{\scriptscriptstyle {\rm ext}}}^{\ast}$,
as well as their respective self-energy renormalization on the Nambu
off diagonal, act as a source and sink for electron pairs, respectively.
Consequently, these terms alter the conservation laws obeyed by the
exact system as compared to the normal state system. Based on the
Dyson Eq.~(\ref{eq:GFDysonEquation}) in the form $\bar{G}^{-1}({\it {\it 1}},{\it 2})=\bar{G}_{{\rm {\scriptscriptstyle H}}}^{-1}({\it 1},{\it 2})-\bar{\varSigma}({\it 1},{\it 2})$,
we generalize the approach of Baym and Kadanoff \citep{BaymConservationLawsAndCorrelationFunctions1961}
to superconductors. In particular, we derive the continuity equation
for the electronic charge. First, we realize 
\begin{eqnarray}
\updelta_{{\it 1},{\it 2}} & = & \bar{G}({\it {\it 1}},{\it 3})\cdot\bar{G}_{{\rm {\scriptscriptstyle 0}}}^{-1}({\it 3},{\it 2})\nonumber \\
 & - & \bar{G}({\it {\it 1}},{\it 3})\cdot\bigl(\bar{\varSigma}({\it 3},{\it 2})-\varPhi({\it 3})\sigma_{0}\tau_{z}\delta_{{\it 3},{\it 2}}\bigr)\label{eq:ContinuityDerivationPart1}\\
\updelta_{{\it 1},{\it 2}} & = & \bar{G}_{{\rm {\scriptscriptstyle 0}}}^{-1}({\it 1},{\it 3})\cdot\bar{G}({\it {\it 3}},{\it 2})\nonumber \\
 & - & \bigl(\bar{\varSigma}({\it 1},{\it 3})-\varPhi({\it 1})\sigma_{0}\tau_{z}\delta_{{\it 1},{\it 3}}\bigr)\cdot\bar{G}({\it {\it 3}},{\it 2})\label{eq:ContinuityDerivationPart2}
\end{eqnarray}
where $\bigl(-\tau_{0}\sigma_{0}\partial_{\tau_{1}}\updelta_{{\it 1},{\it 3}}-\hat{\bar{H}}_{{\rm {\scriptscriptstyle 0}}}({\it 1},{\it 3})\bigr)\bar{G}_{{\rm {\scriptscriptstyle 0}}}({\it 3},{\it 2})=\updelta_{{\it 1},{\it 2}}$
and thus
\begin{eqnarray}
\bar{G}_{{\rm {\scriptscriptstyle 0}}}^{-1}({\it 1},{\it 2}) & = & \bigl(-\tau_{0}\sigma_{0}\partial_{\tau_{1}}\updelta_{{\it 1},{\it 2}}-\hat{\bar{H}}_{{\rm {\scriptscriptstyle 0}}}({\it 1},{\it 2})\bigr)
\end{eqnarray}
Noting that the derivative is anti-symmetric, i.~e.~$\partial_{\tau_{1}}\updelta_{{\it 1},{\it 2}}=-\partial_{\tau_{2}}\updelta_{{\it 1},{\it 2}}$
and $\boldsymbol{\nabla}_{\boldsymbol{r}_{1}}\updelta_{{\it 1},{\it 2}}=-\boldsymbol{\nabla}_{\boldsymbol{r}_{1}}\updelta_{{\it 1},{\it 2}}$,
we calculate 
\begin{eqnarray}
 &  & \int{\rm d}{\it 3}\bigl(\bar{G}_{{\rm {\scriptscriptstyle 0}}}^{-1}({\it 1},{\it 3})\cdot\bar{G}({\it {\it 3}},{\it 2})-\bar{G}({\it {\it 1}},{\it 3})\cdot\bar{G}_{{\rm {\scriptscriptstyle 0}}}^{-1}({\it 3},{\it 2})\bigr)=\nonumber \\
 & = & \int{\rm d}{\it 3}\bigl(\bar{\varSigma}({\it 1},{\it 3})-\varPhi({\it 1})\sigma_{0}\tau_{z}\delta_{{\it 1},{\it 3}}\bigr)\cdot\bar{G}({\it {\it 3}},{\it 2})\nonumber \\
 & - & \bar{G}({\it {\it 1}},{\it 3})\cdot\bigl(\bar{\varSigma}({\it 3},{\it 2})-\varPhi({\it 3})\sigma_{0}\tau_{z}\delta_{{\it 3},{\it 2}}\bigr)\label{eq:SelfEnergyConditionFromDyson}
\end{eqnarray}
in the Appendix \ref{sec:Conservation-conditions}. Similar to Baym
and Kadanoff \citep{BaymConservationLawsAndCorrelationFunctions1961},
in addition taking into account the presence of the SC condensate
many conservation laws can be deduced from this equation. We are particularly
interested in the local ${\it 2}\rightarrow{\it 1}$ limit of the
$1,1$ Nambu component of Eq.~(\ref{eq:G0PartOfTheSelfEnergyConditions}).
If we take the trace in spin space, we arrive at the continuity equation
for the electric charge in a SC
\begin{eqnarray}
 &  & {\rm i}\partial_{\tau_{1}}\langle\hat{n}({\it 1})\rangle+\boldsymbol{\nabla}_{\boldsymbol{r}_{1}}\cdot\langle\hat{\boldsymbol{j}}({\it 1})\rangle=\nonumber \\
 & = & \frac{1}{2}\Im\Bigl(\boldsymbol{\varSigma}_{\Delta}^{\mathfrak{R}}({\it 1},{\it 3})\cdot{\rm Tr}_{{\scriptscriptstyle {\rm s}}}\{\boldsymbol{\Phi}\bar{G}({\it {\it 3}},-1,{\it 1},1)\}\Bigr)\nonumber \\
 & - & \frac{{\rm i}}{2}\Re\Bigl(\boldsymbol{\varSigma}_{\Delta}^{\mathfrak{I}}({\it 1},{\it 3})\cdot{\rm Tr}_{{\scriptscriptstyle {\rm s}}}\{\boldsymbol{\Phi}\bar{G}({\it 3},-1,{\it 1},1)\}\Bigr)\,.\label{eq:ChargeContinuityEquation}
\end{eqnarray}
Here, we have used
\begin{eqnarray}
\langle\hat{n}({\it 1})\rangle & = & {\rm Tr}_{{\scriptscriptstyle {\rm s}}}\{\sigma_{0}\bar{G}({\it {\it 1}},1,{\it 1},1)\}\\
\langle\hat{\boldsymbol{j}}({\it 1})\rangle & = & {\rm Tr}_{{\scriptscriptstyle {\rm s}}}\{\sigma_{0}\bigl(\frac{1}{2{\rm i}}(\boldsymbol{\nabla}_{\boldsymbol{r}_{1}}-\boldsymbol{\nabla}_{\boldsymbol{r}_{2}})+\nonumber \\
 & + & \boldsymbol{A}_{{\scriptscriptstyle {\rm ext}}}({\it 1})\bigr)\bar{G}({\it {\it 1}},1,{\it 2},1)\vert_{{\it 2}\rightarrow{\it 1}}\}
\end{eqnarray}
The right hand side of Eq.~(\ref{eq:ChargeContinuityEquation}) requires
some further explanation. To arrive at Eq.~(\ref{eq:ChargeContinuityEquation})
we have used that $\bar{G}({\it {\it 3}},-1,{\it 1},1)=\bigl(\bar{G}({\it 1},1,{\it {\it 3}},-1)^{{\rm T}_{{\scriptscriptstyle {\rm s}}}}\bigr)^{\ast}$
and introduced the four component vector ($i=1,2,3,4$) 
\begin{eqnarray}
{\varSigma_{\Delta}^{\mathfrak{R}}}_{i}({\it 1},{\it 2}) & = & \frac{1}{2}{\rm Tr}_{{\scriptscriptstyle {\rm s}}}\bigl\{\Phi_{i}^{{\rm T}_{{\rm {\scriptscriptstyle s}}}}\mathfrak{Re}\{\bar{\varSigma}({\it 1},1,{\it 2},-1)\}\bigr\}\nonumber \\
 & - & {\varDelta_{{\scriptscriptstyle {\rm ext}}}}_{i}({\it 1},{\it 2})\\
{\varSigma_{\Delta}^{\mathfrak{I}}}_{i}({\it 1},{\it 2}) & = & \frac{1}{2}{\rm Tr}_{{\scriptscriptstyle {\rm s}}}\bigl\{\Phi_{i}^{{\rm T}_{{\rm {\scriptscriptstyle s}}}}\mathfrak{Im}\{\bar{\varSigma}({\it 1},1,{\it 2},-1)\}\bigr\}\,.
\end{eqnarray}
In this relation, we use the adjoined of a time and space and spin
and Nambu matrix as $\bigl(\bar{\varSigma}({\it 1},{\it 2})\bigr)^{\dagger}=\bigl(\bar{\varSigma}({\it 2},{\it 1})^{\ast}\bigr)^{{\rm T}_{{\scriptscriptstyle {\rm ns}}}}$.
Thus, we see that all source terms for electronic charge depend explicitly
on the SC Nambu off-diagonal terms and vanish in the normal state
as expected. The self-energy is not hermitian in general. We may,
however, decompose it into hermitian and anti hermitian parts by
\begin{eqnarray}
\mathfrak{Re}\{\bar{\varSigma}({\it 1},{\it 2})\} & = & \frac{1}{2}\bigl(\bar{\varSigma}({\it 1},{\it 2})+\bar{\varSigma}({\it 1},{\it 2})^{\dagger}\bigr)\\
\mathfrak{Im}\{\bar{\varSigma}({\it 1},{\it 2})\} & = & \frac{1}{2}\bigl(\bar{\varSigma}({\it 1},{\it 2})-\bar{\varSigma}({\it 1},{\it 2})^{\dagger}\bigr)
\end{eqnarray}
and thus by definition
\begin{eqnarray}
\mathfrak{Re}\{\bar{\varSigma}({\it 1},-1,{\it 2},1)\} & = & \Bigl(\bigl(\mathfrak{Re}\{\bar{\varSigma}({\it 2},1,{\it 1},-1)\}\bigr)^{{\rm T}_{{\scriptscriptstyle {\rm s}}}}\Bigr)^{\ast}\\
\mathfrak{Im}\{\bar{\varSigma}({\it 1},-1,{\it 2},1)\} & = & -\Bigl(\bigl(\mathfrak{Im}\{\bar{\varSigma}({\it 2},1,{\it 1},-1)\}\bigr)^{{\rm T}_{{\scriptscriptstyle {\rm s}}}}\Bigr)^{\ast}\,.
\end{eqnarray}
The four component vector $\boldsymbol{\varSigma}_{\Delta}^{\mathfrak{R}}$
($\boldsymbol{\varSigma}_{\Delta}^{\mathfrak{I}}$) is a spin symmetrized
composition of the (anti) hermitian self-energy and external pair
potential. Any spin matrix can be decomposed into spin symmetric and
antisymmetric components according to $A_{\mu\mu^{\prime}}={\rm Tr}_{{\scriptscriptstyle {\rm s}}}\{\boldsymbol{\Phi}^{{\rm T}_{{\rm {\scriptscriptstyle s}}}}A\}/2\cdot\boldsymbol{\Phi}_{\mu\mu^{\prime}}$.
This is because the Pauli matrices are traceless but $\sigma_{i}^{2}=\sigma_{0}$
so that ${\rm Tr}_{{\scriptscriptstyle {\rm s}}}\{\sigma_{i}A\}$
is a projection on the $\sigma_{i}$ axis of a spin matrix $A$. 
Thus ${\rm Tr}_{{\scriptscriptstyle {\rm s}}}\{\boldsymbol{\Phi}^{{\rm T}_{{\rm {\scriptscriptstyle s}}}}\bar{\varSigma}\}/2$
symmetrizes the superconducting Nambu off diagonal part of the self-energy
into spin anti symmetric ($\Phi_{1}$) and symmetric ($\Phi_{2},\Phi_{3},\Phi_{4}$)
parts. In a static scenario $\partial_{\tau_{1}}\langle\hat{n}({\it 1})\rangle=0$
while in the time dependent case Eq.~(\ref{eq:ChargeContinuityEquation})
remains essentially unchanged except that ${\rm i}\partial/\partial\tau_{1}\rightarrow\partial/\partial z$
where $z$ is on the Keldysh contour \citep{StefanucciNonequilibriumMBTOfQuantumSystemsAModernIntroduction2013}.

\section{The effective interaction\label{sec:TheEffectiveInteraction}}

\begin{figure}
\begin{centering}
\includegraphics[width=1\columnwidth]{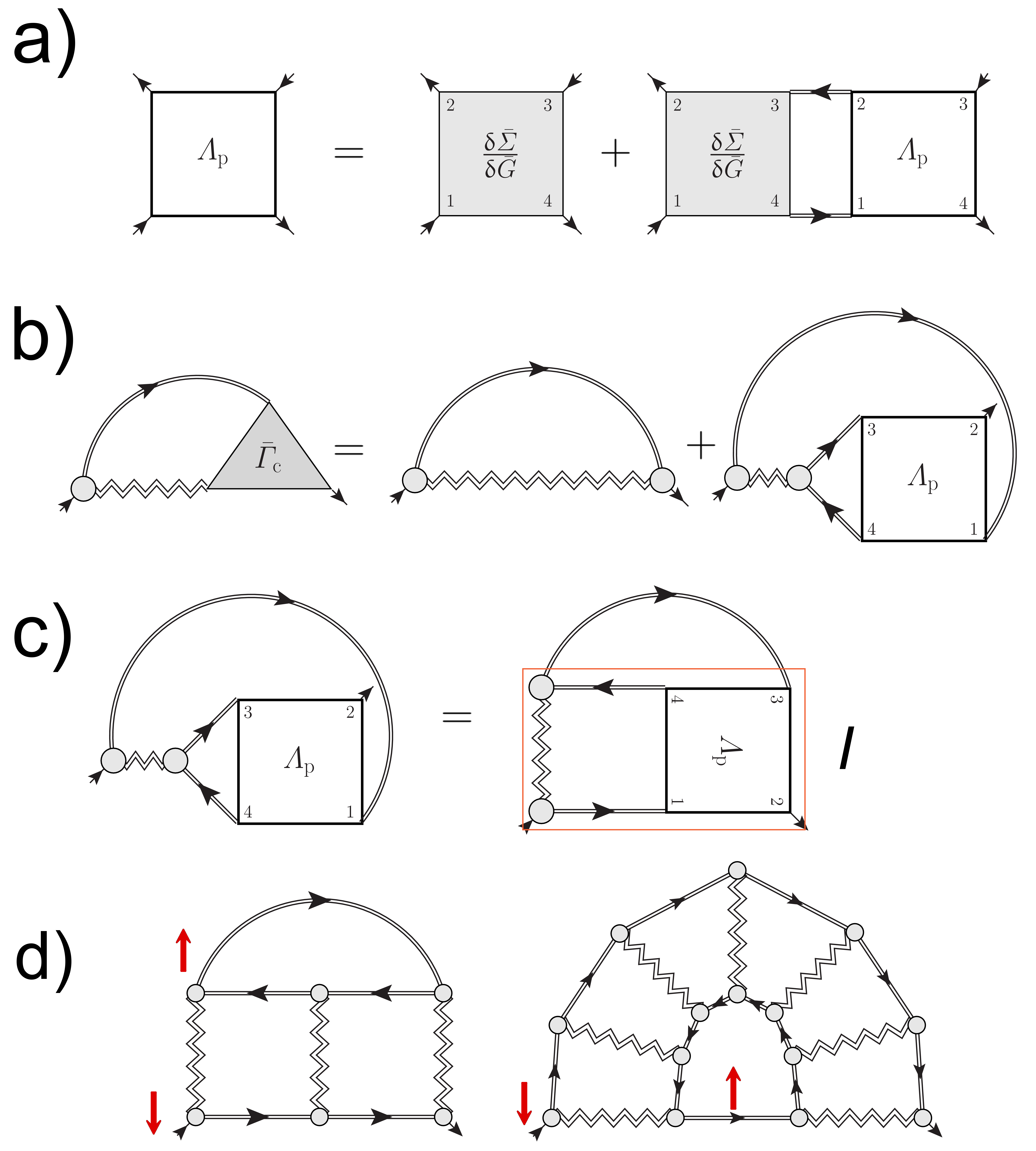}
\par\end{centering}

\caption{a) Diagrammatic representation of the Dyson equation for the SC particle-hole
propagator. b) The electronic self-energy in terms of the particle-hole
propagator. c) Interpretation of the vertex part as an exchange diagram
with an effective interaction as indicated by the box. d) Contributions
to the self-energy from the effective interaction $I$. The arrows
indicate that the electrons may have different spin.\label{fig:DiagramaticRewriting}}
\end{figure}

Among the most important questions in the context of unconventional
SC is where spin and other electronic fluctuations appear in the formalism.
The wording fluctuations is often used meaning beyond mean field theory.
In mean field theory any two particle operator is replaced with a
single particle operator that couples to the average in the resulting
approximate system of the respective other particle. In our case,
where the electrons interact via the Coulomb potential and with a
phonon field, the mean field theory is Hartree-Fock. Similarly, BCS
theory \citep{BCS1957} is a mean field theory \citep{NambuQParticlesGaugeInSuperconductivity1960},
however not on the level of the Coulomb interaction but on the level
of an effective interaction that couples single particle states. This
effective interaction may be based on e.g.~spin flip processes which
cannot be described in a mean field theory of paramagnetic electrons
that interaction via a Coulomb potential. As noted in the introduction
Sec.~\ref{sec:Introduction} we distinguish fluctuation contributions
in our self-energy as the contributions beyond the screened Coulomb
diagram. The reason is that the unscreened Coulomb potential $w_{{\rm {\scriptscriptstyle 0}}}$
does not appear in the self-energy and is ``very far'' from the
screened potential in a metal. Thus, first, we separate the screened
Coulomb diagram from the rest of the electronic self-energy and continue
the discussion based on the terms beyond the screened Coulomb diagram.
We generalize the approach of Essenberger \textit{et al.}~\citep{EssenbergerSCPairingMediatedBySpinFluctuationsFromFirstPrinciples2015}
and introduce the particle-hole propagator $\varLambda_{{\rm {\scriptscriptstyle p}}}$
for a SC by
\begin{eqnarray}
\bar{\varGamma}_{{\scriptscriptstyle {\rm C}}}({\it 1},{\it 2};{\it 3}) & = & \bar{\varGamma}_{{\scriptscriptstyle {\rm C}}}^{{\scriptscriptstyle {\rm 0}}}({\it 1},{\it 2};{\it 3})+\varLambda_{{\rm {\scriptscriptstyle p}}}({\it 1}\bar{{\it 1}},{\it 2}\bar{{\it 2}},{\it 4}\bar{{\it 4}},{\it 5}\bar{{\it 5}})\times\nonumber \\
 & \times & \bar{G}({\it 4}\bar{{\it 4}},{\it 6}\bar{{\it 6}})\bar{\varGamma}_{{\scriptscriptstyle {\rm C}}}^{{\scriptscriptstyle {\rm 0}}}({\it 6}\bar{{\it 6}},{\it 7}\bar{{\it 7}};{\it 3})\bar{G}({\it 7}\bar{{\it 7}},{\it 5}\bar{{\it 5}})\,,\nonumber \\
\label{eq:ElectronicVertexInTermsOfPHProp}
\end{eqnarray}
where $\varLambda_{{\rm {\scriptscriptstyle p}}}$ satisfies the Dyson
equation
\begin{eqnarray}
\varLambda_{{\rm {\scriptscriptstyle p}}}({\it 1}\bar{{\it 1}},{\it 2}\bar{{\it 2}},{\it 3}\bar{{\it 3}},{\it 4}\bar{{\it 4}}) & = & \frac{\updelta\bar{\varSigma}({\it 1}\bar{{\it 1}},{\it 2}\bar{{\it 2}})}{\updelta\bar{G}({\it 3}\bar{{\it 3}},{\it 4}\bar{{\it 4}})}+\frac{\updelta\bar{\varSigma}({\it 1}\bar{{\it 1}},{\it 2}\bar{{\it 2}})}{\updelta\bar{G}({\it 5}\bar{{\it 5}},{\it 6}\bar{{\it 6}})}\times\nonumber \\
 & \times & \bar{G}({\it 5}\bar{{\it 5}},{\it 7}\bar{{\it 7}})\bar{G}({\it 8}\bar{{\it 8}},{\it 6}\bar{{\it 6}})\times\nonumber \\
 & \times & \varLambda_{{\rm {\scriptscriptstyle p}}}({\it 7}\bar{{\it 7}},{\it 8}\bar{{\it 8}},{\it 3}\bar{{\it 3}},{\it 4}\bar{{\it 4}})\,.\label{eq:DysonEquationParticleHolePropagator}
\end{eqnarray}
This may be written diagrammatically as given in Fig.~\ref{fig:DiagramaticRewriting}
a).
Inserting Eq.~(\ref{eq:ElectronicVertexInTermsOfPHProp}) into the
electronic self-energy Eq.~(\ref{eq:ElectronicSelfEnergy}) our separation
of higher vertex corrections in terms of the particle-hole propagator
is complete. The procedure is represented using diagrams as given
in Fig.~\ref{fig:DiagramaticRewriting} b). As an important fact
we realize that the screened interaction $w({\it 1},{\it 2})$ Eq.~(\ref{eq:screendDynamicCoulomb}),
similar to the bare Coulomb interaction in Hartree-Fock, is independent
on spin and the bare vertex conserves angular momentum. Thus, the
first self-energy diagram in Fig.~\ref{fig:DiagramaticRewriting}
b) cannot account for spin flip processes that are discussed as the
possible pairing mechanism in the context of High-$T_{{\rm {\scriptscriptstyle c}}}$
superconductivity. The screened Coulomb exchange diagram will, however,
contain part of the charge-fluctuations in the sense of the fluctuation-exchange
approximation \citep{BickersConservingApproximationForStronglyCorrelatedElSystems1989}.
We conclude that angular momentum transfer processes are described
by the higher order vertex contributions, i.e.~the second diagram
of Fig.~\ref{fig:DiagramaticRewriting} b). This second self-energy
contribution can be twisted to look like an exchange diagram in terms
of an effective interaction as indicated by the thin box in Fig.~\ref{fig:DiagramaticRewriting}
c). In the following we refer to the effective interaction in the
box as $I({\it 1}\bar{{\it 1}},{\it 2}\bar{{\it 2}},{\it 3}\bar{{\it 3}},{\it 4}\bar{{\it 4}})$
where we have labeled the degrees of freedom ${\it 1}\bar{{\it 1}},{\it 2}\bar{{\it 2}},{\it 3}\bar{{\it 3}},{\it 4}\bar{{\it 4}}$
starting with ${\it 1}\bar{{\it 1}}$ at the left bottom of the box
and then continuing clockwise.

This effective interaction has very interesting properties. For example
it can propagate angular momentum and other quantities that are conserved
at the bare vertex due to underlying symmetries. This is possible
since the diagrams that constitute the effective interaction contain
a subset that has no Green's function connection between ``top''
(${\it 2}\bar{{\it 2}},{\it 3}\bar{{\it 3}}$) and ``bottom'' (${\it 1}\bar{{\it 1}},{\it 4}\bar{{\it 4}}$).

An example of such contributions are the horizontal ladder diagrams
beyond second order. First and second order are part of the screened
Coulomb self-energy and thus have to be excluded to avoid double counting.
In this type of diagrams, the conservation at the bare vertex appears
independently on the connected top $({\it 2}\bar{{\it 2}}\leftarrow{\it 3}\bar{{\it 3}})$
and bottom (${\it 1}\bar{{\it 1}}\rightarrow{\it 4}\bar{{\it 4}}$)
Green's function lines. Instead, top and bottom are connected only
with interaction lines and thus the diagram, viewed as a whole, does
not obey the conservation constrains of the bare vertex. Examples
of such contributions are given in Fig.~\ref{fig:DiagramaticRewriting}
d). For the second contribution we have chosen the diagrammatic arrangement
of Doniach and Engelsberg \citep{DoniachLowTemperaturePropertiesOfNearlyFerromagneticFermiLiquids1966}.

For example, a paramagnetic system subject to a Coulomb interaction
has a bare vertex that is proportional to $\sigma_{0}$ in spin space.
Still, this type of effective interaction in the box of Fig.~\ref{fig:DiagramaticRewriting}
c) can propagate angular momentum in the sense that it has components
proportional to $\sigma_{x,y,z}$ as well, i.e.~it describes spin
flip processes. Similarly, it also allows to describe how off diagonal
Nambu components are rotated due to the interaction. The bare electron-electron
and electron-phonon vertex in Nambu space is proportional to $\tau_{z}$
instead of $\sigma_{0}$ as for the spin space, due to the fermionic
commutation rules. The box of Fig.~\ref{fig:DiagramaticRewriting}
c) corresponds to an interaction that has also components in $\tau_{x,y}$
and thus in the language of Anderson \citep{Anderson1958RPAInTheoryOfSuperconductivity}
propagates a rotation of iso-spin. These components correspond to
the Cooper channel and an interaction mediated by the fluctuation
propagator in the sense of Ref.~\onlinecite{LarkinFluctuationPhenomenaInSuperconductors2008}.

$I({\it 1}\bar{{\it 1}},{\it 2}\bar{{\it 2}},{\it 3}\bar{{\it 3}},{\it 4}\bar{{\it 4}})$
is a function of four Nambu and spin indices and time and space variables,
respectively. We suppress the time and space arguments and decompose
$I(\bar{{\it 1}},\bar{{\it 2}},\bar{{\it 3}},\bar{{\it 4}})$ along
the Nambu and spin basis
\begin{eqnarray}
I(\bar{{\it 1}},\bar{{\it 2}},\bar{{\it 3}},\bar{{\it 4}}) & = & \sum_{ijab}I_{ij}^{ab}(\sigma_{a}\tau_{i}\otimes\sigma_{b}\tau_{j})_{\bar{{\it 1}}\bar{{\it 2}}\bar{{\it 3}}\bar{{\it 4}}}\\
I_{ij}^{ab} & = & \sum_{\bar{{\it 1}}\bar{{\it 2}}\bar{{\it 3}}\bar{{\it 4}}}I(\bar{{\it 1}},\bar{{\it 2}},\bar{{\it 3}},\bar{{\it 4}})(\sigma_{a}\tau_{i})_{\bar{{\it 1}}\bar{{\it 2}}}(\sigma_{b}\tau_{j})_{\bar{{\it 3}}\bar{{\it 4}}}\,.
\end{eqnarray}
We write $(\tau_{i}\otimes\tau_{j})_{\bar{{\it 1}}\bar{{\it 2}}\bar{{\it 3}}\bar{{\it 4}}}=(\tau_{i})_{\bar{{\it 1}}\bar{{\it 2}}}(\tau_{j})_{\bar{{\it 3}}\bar{{\it 4}}}$
to indicate that the basis vectors $\tau_{i}$ and $\tau_{j}$ form
an outer product. In the following we discuss only the Nambu degrees
of freedom since the spin is completely analogue. Considering the
Nambu $x,y$ symmetric part of the effective interaction, we may further
decompose 
\begin{eqnarray}
\sum_{ij=x,y}I_{ij}\tau_{i}\otimes\tau_{j} & = & I_{-+}(\tau_{x}-{\rm i}\tau_{y})\otimes(\tau_{x}+{\rm i}\tau_{y})+\nonumber \\
 &  & +I_{+-}(\tau_{x}+{\rm i}\tau_{y})\otimes(\tau_{x}-{\rm i}\tau_{y})+\nonumber \\
 &  & +I_{++}(\tau_{x}+{\rm i}\tau_{y})\otimes(\tau_{x}+{\rm i}\tau_{y})+\nonumber \\
 &  & +I_{--}(\tau_{x}-{\rm i}\tau_{y})\otimes(\tau_{x}-{\rm i}\tau_{y})\bigr)
\end{eqnarray}
with
\begin{eqnarray}
I_{-+} & = & (I_{xx}+I_{yy}-{\rm i}I_{xy}+{\rm i}I_{yx})/4\\
I_{+-} & = & (I_{xx}+I_{yy}+{\rm i}I_{xy}-{\rm i}I_{yx})/4\\
I_{++} & = & (I_{xx}-I_{yy}+{\rm i}I_{xy}+{\rm i}I_{yx})/4\\
I_{--} & = & (I_{xx}-I_{yy}-{\rm i}I_{xy}-{\rm i}I_{yx})/4
\end{eqnarray}
Note that for any complex number $Z$ we find
\begin{eqnarray}
(\tau_{x}\pm{\rm i}\tau_{y})Z & = & {\rm sign(}\Re Z)\vert Z\vert(\tau_{x}\pm{\rm i}\tau_{y}){\rm e}^{\mp{\rm i}\tau_{z}\phi}\\
(\tau_{y}\pm{\rm i}\tau_{z})Z & = & {\rm sign(}\Re Z)\vert Z\vert(\tau_{y}\pm{\rm i}\tau_{z}){\rm e}^{\mp{\rm i}\tau_{x}\phi}\\
(\tau_{z}\pm{\rm i}\tau_{x})Z & = & {\rm sign(}\Re Z)\vert Z\vert(\tau_{z}\pm{\rm i}\tau_{x}){\rm e}^{\mp{\rm i}\tau_{y}\phi}
\end{eqnarray}
where $\phi=\text{arctan}(\Im Z/\Re Z)$. Also note that$\tau_{x}{\rm e}^{\mp{\rm i}\tau_{z,y}\phi}={\rm e}^{\pm{\rm i}\tau_{z,y}\phi}\tau_{x}$
and similar for the other Pauli matrices. Thus, we may write 
\begin{eqnarray}
(\tau_{x}\pm{\rm i}\tau_{y})\otimes(\tau_{x}\pm{\rm i}\tau_{y})I_{\pm\pm} & = & {\rm sign(}\Re I_{\pm\pm})\vert I_{\pm\pm}\vert\times\nonumber \\
 & \times & (\tau_{x}\pm{\rm i}\tau_{y}){\rm e}^{\mp{\rm i}\tau_{z}\frac{\phi_{\pm\pm}}{2}}\otimes\nonumber \\
 & \otimes & {\rm e}^{\pm{\rm i}\tau_{z}\frac{\phi_{\pm\pm}}{2}}(\tau_{x}\pm{\rm i}\tau_{y})
\end{eqnarray}
where
\begin{eqnarray}
\phi_{\pm\pm} & = & \text{arctan}(\Im I_{\pm\pm}/\Re I_{\pm\pm})\,.
\end{eqnarray}
Since the effective interaction $I$ is used in an exchange diagram,
we note that $(\tau_{x}\pm{\rm i}\tau_{y})\cdot\bar{G}\cdot(\tau_{x}\pm{\rm i}\tau_{y})$
interchanges the $(\pm1,\mp1)$ component of the Green function with
the $(\mp1,\pm1)$ component in Nambu space. The $(\pm1,\mp1)$ component
is the complex conjugate of the $(\mp1,\pm1)$ component. Such terms
$I_{\pm\pm}$, constructed from $I_{xy}^{ab},I_{yx}^{ab},I_{yy}^{ab}$
and $I_{xx}^{ab}$, are the SC analog of a spin flip interaction such
as the one of Essenberger \textit{et al.}~\citep{EssenbergerSCPairingMediatedBySpinFluctuationsFromFirstPrinciples2015}.
Furthermore, ${\rm e}^{\mp{\rm i}\tau_{z}\frac{\phi_{\pm\pm}}{2}}\cdot\bar{G}\cdot{\rm e}^{\pm{\rm i}\tau_{z}\frac{\phi_{\pm\pm}}{2}}$
corresponds to a phase rotation of the Nambu off diagonal parts by
$\pm\phi_{\pm\pm}$ . Thus, this part of the effective interaction
corresponds to a (non-local) phase transformation of the condensate
where the phase is rotated by $\pm\phi_{\pm\pm}$ and then conjugated.

In our notation, the spin flip interaction corresponds to the terms
$I_{zz}^{xy},I_{zz}^{yx},I_{zz}^{yy}$ and $I_{zz}^{xx}$ since they
correspond similarly to a self-energy contribution $\sim(\sigma_{x}\pm{\rm i}\sigma_{y})\cdot\bar{G}\cdot(\sigma_{x}\pm{\rm i}\sigma_{y})$.

\section{Summary}

We have derived Hedin's equations for superconducting electrons that
are coupled to a system of phonons. Our coupled equations are formally
exact. Following the approach of Baym and Kadanoff, we derive the
continuity equation for a SC where the external pair potential and
its self-energy renormalization appear as source and sink terms for
electronic charge. We point out how we can define an effective interaction
that can describe fluctuations beyond the screened Coulomb diagram,
e.g.~of spin of the electrons and phase of condensate. This interaction
does not suffer from double counting problems and is given rigorously
in terms of the particle-hole propagator.

\paragraph*{Acknowledgment}

We would like express special thanks to F.~Tandetzky and A.~Sanna
for stimulating discussion and A.~Sanna for a careful proof reading
of the manuscript.

\appendix

\section{The Electronic Dyson Equation\label{sec:The-Electronic-Dyson}}

In this Appendix, we complete the derivation of the Dyson equation
Eq.~(\ref{eq:GFDysonEquation}). We collect the contributions to
the Dyson equation of the single particle parts $\hat{H}_{{\rm {\scriptscriptstyle e}}}+\hat{H}_{{\rm {\scriptscriptstyle aux}}}$
of the total Hamiltonian Eq.~(\ref{eq:TotalHamiltonian}) on the
left hand side of the equation of motion Eq.~(\ref{eq:GFEquationOfMotion}).
The commutators ${\rm T}[\hat{H}_{{\scriptscriptstyle {\rm e}-{\rm e}}}+\hat{H}_{{\scriptscriptstyle {\rm e}-{\rm p}}},\hat{\psi}({\it 1}\mu)]$
and ${\rm T}[\hat{H}_{{\scriptscriptstyle {\rm e}-{\rm e}}}+\hat{H}_{{\scriptscriptstyle {\rm e}-{\rm p}}},\hat{\psi}^{\dagger}({\it 1}\mu)]$
are straight forward to evaluate component-wise while the time ordering
symbol is essential to cast the result back into the unified notation
with the Nambu operators $\hat{\varPsi}({\it 1}\bar{{\it 1}})$. We
obtain
\begin{eqnarray}
 &  & \sum_{\bar{{\it 3}}}\int\hspace{-0.2cm}{\rm d}{\it 3}\Bigl(\delta_{{\it 1},{\it 3}}\bigl(\tau_{0}\sigma_{0}\partial_{\tau_{1}}+\sigma_{0}\tau_{z}\varphi({\it 1})\bigl)\hat{\bar{H}}_{{\scriptscriptstyle 0}}({\it 1},{\it 3})\Bigr)_{\bar{{\it 1}}\bar{{\it 3}}}\times\nonumber \\
 & \times & \bar{G}({\it {\it 3}}\bar{{\it 3}},{\it 2}\bar{{\it 2}})=\nonumber \\
 & = & \frac{1}{2}\sum_{\bar{{\it {\it 3}}}}\int\hspace{-0.2cm}{\rm d}{\it {\it 3}}w_{{\scriptscriptstyle {\rm 0}}}({\it 1},3)(\sigma_{0}\tau_{z})_{\bar{{\it {\it 3}}},\bar{{\it {\it 3}}}}(\sigma_{0}\tau_{z})_{\bar{{\it 1}},\bar{{\it 1}}}\times\nonumber \\
 & \times & \bar{G}^{{\scriptscriptstyle (2)}}({\it {\it 3}}\bar{{\it {\it 3}}},{\it 1}\bar{{\it 1}},{\it {\it 3}}\bar{{\it {\it 3}}},{\it 2}\bar{{\it 2}})\nonumber \\
 & - & \sum_{\bar{{\it {\it 3}}},q,i}(\bar{v}_{i})_{\bar{{\it 1}}\bar{{\it {\it 3}}}}g_{i}(\boldsymbol{{\it r}}_{1}q)\langle\hat{a}_{q}(\tau_{1})\hat{\varPsi}({\it 1}\bar{{\it {\it 3}}})\hat{\varPsi}^{\dagger}({\it 2}\bar{{\it 2}})\rangle_{{\rm {\scriptscriptstyle T}}}\,.\label{eq:AppendixEQOfMotion}
\end{eqnarray}
with the two particle Nambu Green's function $\bar{G}^{{\scriptscriptstyle (2)}}({\it {\it 1}}\bar{{\it {\it 1}}},{\it 2}\bar{{\it 2}},{\it {\it 3}}\bar{{\it {\it 3}}},{\it 4}\bar{{\it 4}})=\langle\hat{\varPsi}^{\dagger}({\it 1}\bar{{\it {\it 1}}})\hat{\varPsi}^{\dagger}({\it 2}\bar{{\it 2}})\hat{\varPsi}({\it 3}\bar{{\it {\it 3}}})\hat{\varPsi}({\it 4}\bar{{\it 4}})\rangle_{{\rm {\scriptscriptstyle T}}}$.
Furthermore we derive the relation
\begin{eqnarray}
 &  & {\rm Tr}_{{\scriptscriptstyle {\rm ns}}}\{\frac{1}{2}\sigma_{0}\tau_{z}\cdot\bar{G}({\it 3},{\it 3})\}\bar{G}({\it 1}\bar{{\it 1}},{\it 2}\bar{{\it 2}})-\frac{\updelta\bar{G}({\it 1}\bar{{\it 1}},{\it 2}\bar{{\it 2}})}{\updelta\varphi({\it 3})}=\nonumber \\
 & = & \sum_{\bar{{\it 3}}}\frac{1}{2}(\sigma_{0}\tau_{z})_{\bar{{\it 3}},\bar{{\it 3}}}\bar{G}^{{\scriptscriptstyle (2)}}({\it 3}\bar{{\it 3}},{\it 1}\bar{{\it 1}},{\it 3}\bar{{\it 3}},{\it 2}\bar{{\it 2}})\,,\label{eq:AppendixDerivative1}
\end{eqnarray}
and similarly
\begin{eqnarray}
 &  & \frac{\updelta\bar{G}({\it 1}\bar{{\it 1}},{\it 2}\bar{{\it 2}})}{\updelta J_{q}(\tau)}-\langle\hat{a}_{q}(\tau_{1})\rangle_{{\rm {\scriptscriptstyle T}}}\bar{G}({\it 1}\bar{{\it 1}},{\it 2}\bar{{\it 2}})=\nonumber \\
 &  & \langle\hat{a}_{q}(\tau_{1})\hat{\varPsi}({\it 1}\bar{{\it 1}})\hat{\varPsi}^{\dagger}({\it 2}\bar{{\it 2}})\rangle_{{\rm {\scriptscriptstyle T}}}\,.\label{eq:AppendixDerivative2}
\end{eqnarray}
Furthermore, for a generic field $A$ 
\begin{equation}
\frac{\updelta\bar{G}({\it 1},{\it 2})}{\updelta A({\it 3})}=-\bar{G}({\it 1},{\it 4})\cdot\frac{\updelta\bar{G}^{-1}({\it 4},{\it 5})}{\updelta A({\it 3})}\cdot\bar{G}({\it 5},{\it 2})\,.\label{eq:AppendixDerivative3}
\end{equation}
At this point we introduce the self energy
\begin{eqnarray}
 &  & \bar{\varSigma}({\it 1},{\it 2})=\nonumber \\
 &  & -w_{{\scriptscriptstyle {\rm 0}}}({\it 1},{\it 3})\tau_{z}\sigma_{0}\cdot\bar{G}({\it 1},{\it 4})\cdot\frac{\updelta\bar{G}^{-1}({\it 4},{\it 2})}{\updelta\varphi({\it 3})}\nonumber \\
 &  & -\bar{\varGamma}_{{\scriptscriptstyle {\rm ph}}}^{{\scriptscriptstyle {\rm 0}}}({\it 1},{\it 3};q\tau_{4})\cdot\bar{G}({\it 3},{\it 4})\cdot\frac{\updelta\bar{G}^{-1}({\it 4},{\it 2})}{\updelta J_{q}(\tau_{4})}+\nonumber \\
 &  & +\bar{\varGamma}_{{\scriptscriptstyle {\rm ph}}}^{{\scriptscriptstyle {\rm 0}}}({\it 1},{\it 3};q\tau_{4})\cdot\updelta_{{\it 2},{\it 3}}\sigma_{0}\tau_{0}\langle\hat{a}_{q}(\tau_{4})\rangle\,.\label{eq:AppendixSelfEnergy}
\end{eqnarray}
Inserting the Eqs.~(\ref{eq:AppendixDerivative1}) and (\ref{eq:AppendixDerivative2})
into Eq.~(\ref{eq:AppendixEQOfMotion}), together with the definition
of the Hartree Hamiltonian $\hat{\bar{H}}_{{\rm {\scriptscriptstyle H}}}({\it 1},{\it 3})$,
we arrive at
\begin{eqnarray}
 &  & \int\hspace{-0.2cm}{\rm d}{\it 3}\bigl(-\delta_{{\it 1},{\it 3}}\tau_{0}\sigma_{0}\partial_{\tau_{1}}-\hat{\bar{H}}_{{\scriptscriptstyle {\rm H}}}({\it 1},{\it 3})\bigr)\cdot\bar{G}({\it 3},{\it 2})\nonumber \\
 &  & =\updelta_{{\it 1}{\it 2}}\updelta_{\bar{{\it 1}}\bar{{\it 2}}}+\int\hspace{-0.2cm}{\rm d}{\it 3}\bar{\varSigma}({\it 1},{\it 3})\cdot\bar{G}({\it 3},{\it 2})\,.
\end{eqnarray}
Now we insert $\bar{G}_{{\scriptscriptstyle {\rm H}}}\bar{G}_{{\scriptscriptstyle {\rm H}}}^{-1}$
and with the Hartree Green's function Eq.~(\ref{eq:HartreeGreensfunction}),
applying $\bar{G}_{{\scriptscriptstyle {\rm H}}}$ from the left,
we arrive at Eq.~(\ref{eq:GFDysonEquation}).

\section{Conservation conditions for the Green Function\label{sec:Conservation-conditions}}

In this Appendix, we want to discuss the conditions implied by the
Dyson equation for the electronic Green function Eq.~(\ref{eq:GFDysonEquation}).
We give the non-local basic equation, that we use to derive the continuity
equation. The procedure is a straight forward adaption of Baym and
Kadanoff\citep{BaymConservationLawsAndCorrelationFunctions1961},
noting that $\bar{G}_{{\rm {\scriptscriptstyle 0}}}^{-1}$ and $\bar{G}$
are $4\times4$ matrices that do not commute. The result for the left
hand side of Eq.~(\ref{eq:SelfEnergyConditionFromDyson}) is\begin{widetext}
\begin{eqnarray}
 &  & \int{\rm d}{\it 3}\bigl(\bar{G}_{{\rm {\scriptscriptstyle 0}}}^{-1}({\it 1},{\it 3})\cdot\bar{G}({\it {\it 3}},{\it 2})-\bar{G}({\it {\it 1}},{\it 3})\cdot\bar{G}_{{\rm {\scriptscriptstyle 0}}}^{-1}({\it 3},{\it 2})\bigr)=\nonumber \\
 & = & \Bigl(-\sigma_{0}\tau_{0}(\partial_{\tau_{1}}+\partial_{\tau_{2}})-\sigma_{0}\tau_{z}\frac{1}{2}\bigl(-{\rm i}(\boldsymbol{\nabla}_{\boldsymbol{r}_{1}}+\boldsymbol{\nabla}_{\boldsymbol{r}_{2}})+\boldsymbol{A}_{{\scriptscriptstyle {\rm ext}}}({\it 1})-\boldsymbol{A}_{{\scriptscriptstyle {\rm ext}}}({\it 2})\bigr)\bigl(-{\rm i}(\boldsymbol{\nabla}_{\boldsymbol{r}_{1}}-\boldsymbol{\nabla}_{\boldsymbol{r}_{2}})+\boldsymbol{A}_{{\scriptscriptstyle {\rm ext}}}({\it 1})+\boldsymbol{A}_{{\scriptscriptstyle {\rm ext}}}({\it 2})\bigr)\nonumber \\
 & + & \sigma_{0}\tau_{z}\bigl(\phi_{{\scriptscriptstyle {\rm ext}}}({\it 1})-\phi_{{\scriptscriptstyle {\rm ext}}}({\it 2})\bigr)+\frac{1}{2}\bigl(\mathbf{S}(\tau_{0}+\tau_{z})-\mathbf{S}^{\ast}(\tau_{0}-\tau_{z})\bigr)\cdot\bigl(\boldsymbol{B}_{{\scriptscriptstyle {\rm ext}}}({\it 1})-\boldsymbol{B}_{{\scriptscriptstyle {\rm ext}}}({\it 2})\bigr)\Bigr)\cdot\bar{G}({\it {\it 1}},{\it 2})\nonumber \\
 & - & \Bigl(\bigl({\rm i}\boldsymbol{\nabla}_{\boldsymbol{r}_{2}}+\boldsymbol{A}_{{\scriptscriptstyle {\rm ext}}}({\it 2})\bigr)^{2}+\phi_{{\scriptscriptstyle {\rm ext}}}({\it 2})\Bigr)[\sigma_{0}\tau_{z},\bar{G}({\it {\it 1}},{\it 2})]-\frac{1}{2}[\bigl(\mathbf{S}(\tau_{0}+\tau_{z})-\mathbf{S}^{\ast}(\tau_{0}-\tau_{z})\bigr)\cdot\boldsymbol{B}_{{\scriptscriptstyle {\rm ext}}}({\it 2}),\bar{G}({\it {\it 1}},{\it 2})]\nonumber \\
 & + & \frac{1}{2}\int{\rm d}{\it 3}\bigl((\tau_{x}-{\rm i}\tau_{y})\boldsymbol{\varDelta}_{{\scriptscriptstyle {\rm ext}}}({\it 1},{\it 3})\cdot\boldsymbol{\Phi}-(\tau_{x}+{\rm i}\tau_{y})\boldsymbol{\varDelta}_{{\scriptscriptstyle {\rm ext}}}^{\ast}({\it 1},{\it 3})\cdot\boldsymbol{\Phi}\bigr)\cdot\bar{G}({\it {\it 3}},{\it 2})\nonumber \\
 & - & \frac{1}{2}\int{\rm d}{\it 3}\bar{G}({\it {\it 1}},{\it 3})\cdot\bigl((\tau_{x}-{\rm i}\tau_{y})\boldsymbol{\varDelta}_{{\scriptscriptstyle {\rm ext}}}({\it 3},{\it 2})\cdot\boldsymbol{\Phi}-(\tau_{x}+{\rm i}\tau_{y})\boldsymbol{\varDelta}_{{\scriptscriptstyle {\rm ext}}}^{\ast}({\it 3},{\it 2})\cdot\boldsymbol{\Phi}\bigr)\label{eq:G0PartOfTheSelfEnergyConditions}
\end{eqnarray}
The Nambu off diagonal right hand side of Eq.~(\ref{eq:SelfEnergyConditionFromDyson})
is then evaluated to be
\begin{eqnarray}
 &  & \frac{1}{2}\int{\rm d}{\it 3}\bigl((\tau_{x}+{\rm i}\tau_{y})\bar{\varSigma}({\it 1},1,{\it 3},-1)+(\tau_{x}-{\rm i}\tau_{y})\bar{\varSigma}({\it 1},-1,{\it 3},1)\bigr)\cdot\bar{G}({\it {\it 3}},{\it 1})\nonumber \\
 & - & \frac{1}{2}\int{\rm d}{\it 3}\bar{G}({\it {\it 1}},{\it 3})\cdot\bigl((\tau_{x}+{\rm i}\tau_{y})\bar{\varSigma}({\it 3},1,{\it 2},-1)+(\tau_{x}-{\rm i}\tau_{y})\bar{\varSigma}({\it 3},-1,{\it 2},1)\bigr)\nonumber \\
 & = & \left(\begin{array}{cc}
\mathfrak{Im}\bigl\{\int{\rm d}{\it 3}\mathfrak{Re}\{\bar{\varSigma}({\it 1},1,{\it 3},-1)\}\bar{G}({\it {\it 3}},-1,{\it 2},1)\bigr\} & \mathfrak{Re}\bigl\{\int{\rm d}{\it 3}\mathfrak{Re}\{\bar{\varSigma}({\it 1},1,{\it 3},-1)\}\bar{G}({\it {\it 3}},-1,{\it 2},-1)\bigr\}\\
\mathfrak{Re}\bigl\{\int{\rm d}{\it 3}\mathfrak{Re}\{\bar{\varSigma}({\it 1},-1,{\it 3},1)\}\bar{G}({\it {\it 3}},1,{\it 2},1)\bigr\} & \mathfrak{Im}\bigl\{\int{\rm d}{\it 3}\mathfrak{Re}\{\bar{\varSigma}({\it 1},-1,{\it 3},1)\}\bar{G}({\it {\it 3}},1,{\it 2},-1)\bigr\}
\end{array}\right)\nonumber \\
 &  & +\left(\begin{array}{cc}
\mathfrak{Re}\bigl\{\int{\rm d}{\it 3}\mathfrak{Im}\{\bar{\varSigma}({\it 1},1,{\it 3},-1)\}\bar{G}({\it {\it 3}},-1,{\it 2},1)\bigr\} & \mathfrak{Im}\bigl\{\int{\rm d}{\it 3}\mathfrak{Im}\{\bar{\varSigma}({\it 1},1,{\it 3},-1)\}\bar{G}({\it {\it 3}},-1,{\it 2},-1)\bigr\}\\
\mathfrak{Im}\bigl\{\int{\rm d}{\it 3}\mathfrak{Im}\{\bar{\varSigma}({\it 1},-1,{\it 3},1)\}\bar{G}({\it {\it 3}},1,{\it 2},1)\bigr\} & \mathfrak{Re}\bigl\{\int{\rm d}{\it 3}\mathfrak{Im}\{\bar{\varSigma}({\it 1},-1,{\it 3},1)\}\bar{G}({\it {\it 3}},1,{\it 2},-1)\bigr\}
\end{array}\right)\,.
\end{eqnarray}
The Nambu diagonal part can be computed in an analogous way. Taking
the local limit ${\it 2}\rightarrow{\it 1}$ of the Eq.~(\ref{eq:G0PartOfTheSelfEnergyConditions})
gives
\begin{eqnarray}
 &  & \int{\rm d}{\it 3}\bigl(\bar{G}_{{\rm {\scriptscriptstyle 0}}}^{-1}({\it 1},{\it 3})\cdot\bar{G}({\it {\it 3}},{\it 1})-\bar{G}({\it {\it 1}},{\it 3})\cdot\bar{G}_{{\rm {\scriptscriptstyle 0}}}^{-1}({\it 3},{\it 1})\bigr)=\nonumber \\
 & = & \Bigl(-2\sigma_{0}\tau_{0}\partial_{\tau_{1}}\bar{G}({\it {\it 1}},{\it 2})\vert_{{\it 2}\rightarrow{\it 1}}+2\sigma_{0}\tau_{z}{\rm i}\boldsymbol{\nabla}_{\boldsymbol{r}_{1}}\bigl(\frac{1}{2{\rm i}}(\boldsymbol{\nabla}_{\boldsymbol{r}_{1}}-\boldsymbol{\nabla}_{\boldsymbol{r}_{2}})+\boldsymbol{A}_{{\scriptscriptstyle {\rm ext}}}({\it 1})\bigr)\bar{G}({\it {\it 1}},{\it 2})\vert_{{\it 2}\rightarrow{\it 1}}\nonumber \\
 & - & \Bigl(\bigl({\rm i}\boldsymbol{\nabla}_{\boldsymbol{r}_{2}}+\boldsymbol{A}_{{\scriptscriptstyle {\rm ext}}}({\it 1})\bigr)^{2}+\phi_{{\scriptscriptstyle {\rm ext}}}({\it 1})\Bigr)[\sigma_{0}\tau_{z},\bar{G}({\it {\it 1}},{\it 2})]\vert_{{\it 2}\rightarrow{\it 1}}-\frac{1}{2}[\bigl(\mathbf{S}(\tau_{0}+\tau_{z})-\mathbf{S}^{\ast}(\tau_{0}-\tau_{z})\bigr)\cdot\boldsymbol{B}_{{\scriptscriptstyle {\rm ext}}}({\it 1}),\bar{G}({\it {\it 1}},{\it 1})]\nonumber \\
 & + & \frac{1}{2}\int{\rm d}{\it 3}\bigl((\tau_{x}+{\rm i}\tau_{y})\boldsymbol{\varDelta}_{{\scriptscriptstyle {\rm ext}}}({\it 1},{\it 3})\cdot\boldsymbol{\Phi}-(\tau_{x}-{\rm i}\tau_{y})\boldsymbol{\varDelta}_{{\scriptscriptstyle {\rm ext}}}^{\ast}({\it 1},{\it 3})\cdot\boldsymbol{\Phi}\bigr)\cdot\bar{G}({\it {\it 3}},{\it 1})\nonumber \\
 & - & \frac{1}{2}\int{\rm d}{\it 3}\bar{G}({\it {\it 1}},{\it 3})\cdot\bigl((\tau_{x}+{\rm i}\tau_{y})\boldsymbol{\varDelta}_{{\scriptscriptstyle {\rm ext}}}({\it 3},{\it 1})\cdot\boldsymbol{\Phi}-(\tau_{x}-{\rm i}\tau_{y})\boldsymbol{\varDelta}_{{\scriptscriptstyle {\rm ext}}}^{\ast}({\it 3},{\it 1})\cdot\boldsymbol{\Phi}\bigr)\,.
\end{eqnarray}
\end{widetext}It is important that the commutator $[\sigma_{0}\tau_{z},\bar{G}({\it {\it 3}},{\it 2})]$
has only components on the Nambu off diagonal. If we are interested
in the usual conservation laws of electronic charge or magnetic density,
these terms do not appear. Taking the local limit and selecting only
the $1,1$ component, the equations simplify significantly and we
arrive at Eq.~(\ref{eq:ChargeContinuityEquation}).

Furthermore, we point out that if the self-energy is non-SC, i.e.~has
only Nambu diagonal non vanishing components$\bar{\varSigma}({\it 1},{\it 3})\propto\tau_{z},\tau_{0}$,
the right hand side of Eq.~(\ref{eq:SelfEnergyConditionFromDyson})
is proportional to the commutator $[\sigma_{0}\tau_{z},\bar{G}({\it {\it 3}},{\it 2})]$.
This commutator on the other hand does not have components on the
Nambu diagonal, i.e.~the usual continuity equations for electronic
charge and magnetic density are satisfied without self-energy contributions
as expected.\bibliographystyle{apsrev4-1}
\bibliography{/home/alinsch/BibTeX/BibTeX}

\end{document}